\documentclass[preprint,12pt]{elsarticle}

\usepackage[utf8]{inputenc}
\usepackage{natbib}
\usepackage{algorithm}
\usepackage[noend]{algpseudocode}
\usepackage[normalem]{ulem}
\usepackage{bm}
\usepackage{dsfont}
\usepackage{hyperref}
\usepackage{xr} 
\usepackage{comment}
\usepackage{amsmath}
\usepackage{amssymb}
\usepackage{mathtools}
\usepackage{graphicx}
\usepackage{graphics}
\usepackage{subfig}
\usepackage{array}
\usepackage{colortbl}
\usepackage{float}
\usepackage{booktabs}
\usepackage{makecell}
\usepackage{multirow}
\usepackage[table]{xcolor}
\usepackage{longtable}

\DeclareMathOperator*{\argmin}{arg\,min}

\newcommand{\one}[1]{\mathds{1}_{#1}}

\algdef{SE}[DOWHILE]{Do}{doWhile}{\algorithmicdo}[1]{\algorithmicwhile\ #1}%

\DeclarePairedDelimiter\abs{\lvert}{\rvert}%
\DeclarePairedDelimiter\norm{\lVert}{\rVert}%
\makeatletter
\let\oldabs\abs
\def\abs{\@ifstar{\oldabs}{\oldabs*}}
\let\oldnorm\norm
\def\norm{\@ifstar{\oldnorm}{\oldnorm*}}
\makeatother

\journal{Computational Statistics and Data Analysis}

\begin{document}

\begin{frontmatter}
\title{Graph-Based Spatial Segmentation of Health-Related Areal Data}

\author[label1,label2,label3,label4]{Vivien Goepp\corref{cor1}}
\author[label5]{Jan van de Kassteele}
\cortext[cor1]{Corresponding author: \texttt{vivien.goepp@gmail.com}\\
CBIO-Centre for Computational Biology, F-75006 Paris, France} 
\address[label1]{Mines ParisTech, PSL Research University, \\ CBIO-Centre for Computational Biology, F-75006 Paris, France}
\address[label2]{Institut Curie, PSL Research University, F-75005 Paris, France}
\address[label3]{INSERM, U900, F-75005 Paris, France}
\address[label4]{MAP5, CNRS UMR 8145, 45, rue des Saints-Pères, 75006, Paris, France}
\address[label5]{National Institute for Public Health and the Environment - RIVM, Bilthoven, \\ The Netherlands}

\begin{abstract}

{\color{black}
Smoothing is often used to improve the readability and interpretability of noisy areal data.
However there are many instances where the underlying quantity is discontinuous. 
In this case, specific methods are needed to estimate the piecewise constant spatial process.
A well-known approach in this setting is to perform segmentation of the signal using the adjacency graph, as does the graph-based fused lasso. But this method does not scale well to large graphs.

This article introduces a new method for piecewise-constant spatial estimation that \emph{(i)} is fast to compute on large graphs and \emph{(ii)} yields sparser models than the fused lasso (for the same amount of regularization), giving estimates that are easier to interpret.

We illustrate our method on simulated data and apply it to real data on overweight prevalence in the Netherlands. Healthy and unhealthy zones are identified which cannot be explained by demographic of socio-economic characteristics. We find that our method is capable of identifying such zones and can assist policy makers with their health-improving strategies.
The implementation of our method in \texttt{R} is publicly available at \url{github.com/goepp/graphseg}.
}
\end{abstract}


\end{frontmatter}


\section{Introduction}

Spatial statistics plays a prominent role in epidemiology. Nowadays, the study of health-related outcomes with respect to the geographical location has become widespread. The data for these studies can be of various types, leading to different statistical tools to answer the epidemiological questions at hand \citep{Lawson2016HandbookSpatialEpidemiology}.

A common problem in spatial statistics is the regularization of spatial data. Most regularization methods perform a spatial smoothing of the data, e.g. kriging for interpolation of data that has been collected at fixed point locations \citep{Cressie1993StatisticsSpatialData} or disease mapping techniques in the case of data has been collected for administrative areas \citep{Lawson2016HandbookSpatialEpidemiology}. The main advantage of smoothing is a higher interpretability of the resulting map. The underlying assumption is that the actual truth is smooth.

In some cases however, one may want to obtain a segmented estimation of the spatial distribution, for instance when the true underlying spatial effect is assumed to be discontinuous. Besides, from a policy making point of view, for the purpose of improving the health of a population, it can be of interest to identify zones having a similar health-related status. Because of logistic and administrative efficiencies involved in such health-improving strategies, the areas within such zones should preferably be contiguous. Spatial segmentation techniques provide an objective way to identify such zones.

In demographic and epidemiological studies, the neighborhood in which we live often has an effect on the health-related outcome variable, even when adjusted for demographic variables, like age and sex, and other socio-economic variables, like educational level and income. This leads to the hypothesis that the neighborhood of residence has by itself an impact on the health-related outcome variable. Beyond the demographic and socio-economic factors, people living in the same area tend to share the same habits: the school they attend, the supermarket they shop at, the bank they go to, etc. As an example, demographic studies on longevity focus namely on finding specific geographic areas where the longevity is unexpectedly high. These more-than-expected healthier areas are sometimes referred to as "blue zones" \citep{Poulain2004Identificationgeographicarea}. These areas are discrete by nature, and such studies use a spatial segmentation of administrative) areas to identify these zones.

Thus, instead of looking at the prevalence of a health-related indicator itself, it may be more interesting for policy makers to identify zones that have a higher or lower prevalence than can be expected based on the demographic and socioeconomic composition of neighborhoods alone.
Such information is usually not directly available, but \citet{vandeKassteele2017Estimatingprevalence26} presented a small area estimation model that, as a by-product, provides an estimate of these neighborhood-specific deviations in the form of a spatial random effect term.
The goal is to identify healthy and unhealthy zones by performing spatial segmentation on the neighborhood specific deviations that cannot be explained by demographic of socio-economic characteristics.

The main approach for segmentation of piecewise spatial data is to use the graph-fused lasso on the adjacency graph. 
The graph-fused lasso was first introduced for regression \citep{Kim2009multivariateregressionapproach} and then extended to multitask regression \citep{Chen2010GraphStructuredMultitaskRegression}.
\citet{Hoefling2010PathAlgorithmFused} introduced a path algorithm for regression with any fused lasso penalty, called generalized fused lasso and \citet{Wang2016TrendFilteringGraphs} developed a method for trend filtering fused lasso on graphs using the ADMM optimization algorithm.
Finally, \cite{Tansey2015FastFlexibleAlgorithm} have proposed a fast estimation procedure for the graph fused lasso with any convex loss based on a trail decomposition of the graph and the ADMM algorithm.

In this paper, we introduce a new graph-based sparsity-inducing estimation method that yields sparser estimates than the graph fused lasso.
As for the graph fused lasso \citep{Hoefling2010PathAlgorithmFused}, the method minimizes a likelihood penalized over the differences of the parameter over a graph, using the adjacency structure of the areas.
Our method extends the adaptive ridge \citep{Rippe2012VisualizationGenomicChanges, Frommlet2016AdaptiveRidgeProcedure} to a graph fused penalty.
The adaptive ridge is a sparsity-inducing iterative method based on a iterating over a weighted ridge problem.
Since the complexity depends on the number of areas and not on the number of individuals, our method is computationally efficient when there is a large number of areas. 

The paper is organized as follows. Section 2.1 introduces the model. In Section 2.2 we investigate the properties of our method, both on simulated data and on real data on overweight prevalence in the Netherlands. Results are shown in Section 3 and are discussed in Section 4.

\section{Methods and data}

\subsection{The graph-based fused adaptive ridge}
In this section, we present our method for estimating the piecewise graph-based signal from a noisy observation.

\subsubsection{Segmentation of spatial lattice data}

The method presented in this section applies to any signal defined on an undirected, simple graph.
However, in this paper, it is applied to the case of lattice spatial data.
We first explain how lattice spatial data can be viewed.

Consider a set of~$p \geq 1$ areas forming a partition of a connected subset of $\mathbb{R}^2$.
In this work, areas often correspond to an administrative division of a territory, for instance census tracts, municipalities, counties, or neighborhoods.
Consider a signal~$\bm{x}\in \mathbb{R}^{p}$ (called \emph{spatial effect} in the following) where each component of~$\bm{x}$ corresponds to an area.
Consider that~$\bm{x}$ is a noisy observation of an unobserved effect $\bm\theta$:
\begin{equation}
  \bm{x} = \bm{\theta} + \bm{\varepsilon},
    \label{eq:model_gaussian}
\end{equation}
where $\bm{\theta} \in \mathbb{R}^{p}$ is the underlying, deterministic signal and~$\bm{\varepsilon}\in \mathbb{R}^{p}$ is an error term centered around zero.

The errors need not be normally distributed. 
When there are, the mean square estimate introduced in this paper corresponds to the negative log-likelihood.

We assume that~$\bm{\theta}$ is piecewise constant.
More precisely, it has the same values on unions of areas, which we call \emph{zones}.
The true number of zones is noted~$q$ and hence~$\bm{\theta}$ only takes~$q$ different values.
The zones form a partition of the areas, and each zone is made of a subset of contiguous areas (rook-type contiguity).
The goal of this paper is to infer~$\bm{\theta}$.

\subsubsection{Spatial lattice data as a signal on graph}

Consider the adjacency graph~$\mathcal{G} = (\mathcal{E}, \mathcal{V})$ of the areas, where each vertex is an area and two vertices are adjacent if their corresponding areas share a border (rook-type contiguity).
We can then view~$\bm{x}$ as a signal on this graph (see \citep{Shuman2013EmergingFieldSignal} for a review of signals on graphs).
Estimating~$\bm{\theta}$ can then be viewed as a problem of piecewise constant estimation of the signal on graph.

Note that when estimating~$\bm{\theta}$ as a signal on graph, we discard any information of the areas as geometrical sets (e.g. Euclidian distance between centroids).
This modelization simplifies the problem, but makes the assumption that adjacency contains sufficently enough information about how \emph{similar} two areas are.

In applications where there are several connected components (e.g. caused by rivers or the presence of islands) one can consider each component as a separate problem, or connect every pair of components by artificially adding an edge between their closest areas (using the Euclidean distance between their centroids).



\subsubsection{Estimation procedure}
Segmentation of $\bm{\theta}$ is done by using a sparsity-inducing method applied to the differences of the values of $\bm\theta$. The penalty method we use is the adaptive ridge \citep{Rippe2012VisualizationGenomicChanges, Frommlet2016AdaptiveRidgeProcedure}. This penalized method belong to the class of sparsity-inducing penalized method, like the lasso \citep{Tibshirani1996RegressionShrinkageSelection}. It  performs feature selection by iterating over re-weighted L$_2$ norm penalties, which have an explicit solution. 

We define the sum-of-squares cost function
\begin{equation}
  \ell(\bm\theta) = \tfrac{1}{2}(\bm x - \bm \theta)^\intercal \bm\Sigma^{-1} (\bm x - \bm \theta),  
  \label{eq:sum_of_squares}
\end{equation}
where~$\bm{\Sigma}$ is the covariance matrix of~$\bm{\varepsilon}$.
As mentioned above, this cost function corresponds to the negative log-likelihood when~$\bm{\varepsilon}\sim \mathcal{N}(\bm{0}, \bm{\Sigma})$.
When the covariance is not assumed to be known, we set~$\bm{\Sigma} = \bm{I}$ in the cost function.

The estimating procedure is as follows.
First, we define the weighted undirected adjacency graph $\mathcal{G} ^{(l)}= (\mathcal{E}, \mathcal{V}^{(l)})$, where each weighted edge $\{j,k\}\in \mathcal{V}^{(l)}$ between vertices $j$ and $k$ is assigned a positive weight $v_{j,k}^{(l)} \geq 0$ (by definition~$j$ and~$k$ are not adjacent if~$v_{j,k}^{(l)} = 0$) that depends on the iteration step~$l$.
Next, define the penalized log-likelihood $\ell^{\text{pen}}$ using a weighted L$_{2}$ penalty:
\begin{equation}
\ell^{\text{pen}}(\bm{\theta}, \mathcal{V}^{(l)}) \triangleq \ell(\bm\theta) + \frac{\lambda}{2} \sum_{j \sim k}v_{j,k}^{(l)} \left(\theta_j - \theta_k\right) ^ 2,
\label{eq:nll_pen}
\end{equation}
where $\lambda > 0$ is a smoothing parameter.
The set of weighted edges $\mathcal{V}^{(l)}$ is included as a parameter of $\ell^{\text{pen}}$ to highlight the dependence on the current weighted graph $(\mathcal{E}, \mathcal{V}^{(l)})$. The sum in the latter equation is taken only once per vertex, that is, the sum index is $\{(j,k) \in \mathcal{E}, j < k\}$, where an arbitrary ordering of the nodes has been chosen. The edge weights $v_{j,k}^{(l)}$ play the role of tuning the importance of the difference between areas $j$ and $k$ while $\lambda$ plays the role of tuning the overall regularization.

The adaptive ridge procedure iterates between a step of weighted smoothing and an update of the weights:
\begin{subequations} \label{eq:ar_algo}
\begin{align}
\text{(i)} \quad & \bm\theta^{(l)} \triangleq \argmin_{\bm\theta} \ell^{\text{pen}}(\bm{\theta}, \mathcal{V}^{(l - 1)}) \label{eq:ar_algo1} \\
\text{(ii)} \quad & v_{j,k}^{(l)} = \frac{1}{(\theta_{j}^{(l)} - \theta_{k}^{(l)}) ^{2} + \varepsilon}, \quad \{j,k\} \in \mathcal{V}.\label{eq:ar_algo2}
\end{align}
\end{subequations}
where $\varepsilon > 0$ is a small numerical constant, introduced in order to bound the denominator away from zero for numerical stability. 
Different choices for the value of $\varepsilon$ have been proposed, \citet{Candes2008EnhancingSparsityReweighted} and \citet{Daubechies2008Iterativelyreweightedleast} having proposed to update its value at each iteration, decreasing it as the algorithm converges.
Numerical experiments \citep{Candes2008EnhancingSparsityReweighted, Frommlet2016AdaptiveRidgeProcedure} have highlighted that the estimation procedure is relatively robust to the choice of $\varepsilon$, so we favored setting a constant $\varepsilon$ ($\varepsilon = 10^{-6}$ in our implementation).
More details about the convergence of Eq.~\eqref{eq:ar_algo} is given in~\ref{sec:appendix_convergence}.

\subsubsection{Implementation and algorithmic considerations}

Define the weighted Laplacian matrix associated to the weighted graph $(\mathcal{E}, \mathcal{V}^{(l)})$: $\bm{K}^{(l)} = \bm{D}^{(l)} - \bm{A}^{(l)}$ where $\bm{D}^{(l)}$ is the diagonal matrix giving the weighted degree of each node: $d_{j,j}^{(l)} = \sum_{k\sim j} v_{j,k}^{(l)}$ and $\bm{A}^{(l)}$ is the weighted adjacency matrix: $a_{j,k}^{(l)} = v_{j,k}^{(l)}$ if $j\sim k$ and zero otherwise.

We first rewrite Eq.~\eqref{eq:nll_pen}. 
Using the fact that $\sum_{j\sim k} v_{j,k}^{(l)} (\theta_j - \theta_k)^2 = \bm\theta^\intercal \bm{K}^{(l)} \bm{\theta}$ and using Eq.~\eqref{eq:sum_of_squares}, the penalized likelihood can be written
\begin{equation}
    \ell^{\text{pen}}(\bm\theta, \mathcal{V}^{(l)}) = \frac{1}{2}(\bm x - \bm \theta)^\intercal \bm\Sigma^{-1} (\bm x - \bm \theta) + \frac{\lambda}{2} \bm\theta^\intercal \bm{K}^{(l)} \bm\theta
    \label{eq:nll_sum_of_squares}
\end{equation}
and the weighted ridge problem in Eq. \eqref{eq:ar_algo} is solved by the explicit update:
\begin{equation}
    \bm\theta^{(l)} = (\bm\Sigma^{-1} + \lambda \bm{K}^{(l - 1)})^{-1}\bm\Sigma ^{-1}\bm x.
    \label{eq:ar_weighted_ridge}
\end{equation}
In the simple case of independent spatial effects, the precision matrix $\bm\Sigma^{-1}$ is diagonal with $j$-th entry $(1/\sigma_j^2)$. If the $x_j$s are not assumed independent, as will be the case in our real data application, we assume that $\bm\Sigma^{-1}$ is sparse. Under this assumption, $\bm\Sigma^{-1} + \lambda \bm K^{(l)}$ is sparse positive definite and its inversion can be done using the Cholesky decomposition. In the application, we use a sparse estimate of $\bm\Sigma^{-1}$.

\paragraph{Remark}
The sum of squares function \label{eq:nll_sum_of_squares} is derived from the likelihood when~$\bm{x}$ is Gaussian.
However we can use the method in the more general setting where~$\bm{x}$ follows any distribution of mean~$\bm{\theta}$ and covariance matrix~$\bm{\Sigma}$.
For computational efficiency, our implementation nonetheless requires that~$\bm{\Sigma}^{-1}$ be sparse.

The computational bottleneck of the iterative procedure is the linear system~\eqref{eq:ar_weighted_ridge}. We use the package \texttt{Matrix}, version 1.2-17, which makes use of the \texttt{CHOLMOD} library \citep{Chen2008Algorithm887CHOLMOD} for fast inversion of sparse (semi-)positive definite matrices. Moreover, the matrix $\bm\Sigma ^{-1} + \lambda \bm K^{(l)}$ has the same sparsity structure at all steps. This can be leveraged to further speed-up the iterative procedure: we compute the symbolic Cholesky decomposition of $\bm\Sigma ^{-1} + \lambda \bm K^{(l)}$ only once (using function \texttt{Matrix::Cholesky}) and update the numerical values at each iteration (using function \texttt{Matrix::update}).

The segmentation procedure for one value of $\lambda$ is given in Algorithm \ref{alg:graph_ar}. In practice, estimation is performed on a grid of penalties, and the choice of the best $\lambda$ is done in a second step.

\begin{algorithm}
\caption{Segmentation over a graph using the adaptive ridge\label{alg:graph_ar}}
\begin{algorithmic}[1]
\Procedure{Adaptive-Ridge}{$\bm x, \bm{\Sigma}^{-1}, \lambda$}
\State $v_{j,k}^{(0)} \gets \one{j \sim k}$
\State $l \gets 1$
\Do
\State $\bm K^{(l - 1)} \gets \bm D^{(l - 1)} - \bm A^{(l - 1)}$
\State $\bm\theta^{(l)} \gets (\bm\Sigma^{-1} + \lambda \bm K^{(l - 1)})^{-1}\bm{\Sigma}^{-1}\bm x$
\State $v_{j, k}^{(l)} \gets ((\theta_j^{(l)} - \theta_k^{(l)})^2 + \varepsilon ) ^{-1}$
\State~$\delta_{j,k}^{(l)} \gets v_{j,k}^{(l)} (\theta_{j}^{(l)} - \theta_{k}^{(l)})^{2}$
\State~$l \gets l + 1$
\doWhile{$\max_{j,k} \abs*{\delta_{j,k}^{(l)} - \delta_{j,k}^{(l - 1)}} > \text{tol}$}
\State \Return{$\bm{\theta}$}
\EndProcedure
\end{algorithmic}
\end{algorithm}

The implementation of the method for a sequence of penalties is made faster using a \emph{warm start}: the estimations are performed on an increasing sequence of penalties $(\lambda_q)$, and the weights $v_{j,k}^{(l)}$ obtained at convergence for $\lambda_q$ are recycled for the first iteration of the estimation $\lambda_{q + 1}$. This trick is based on the fact that the limit of the adaptive ridge iterations does not vary a lot under a small variation in $\lambda$. It significantly reduces the number of iterations needed for convergence for subsequent values of the penalty.

\subsubsection{Choice of the regularization parameter}
In penalized methods, the choice of the penalty $\lambda$ is a difficult task \citep[Section 7]{Hastie2009ElementsStatisticalLearning}. In a number of statistical settings, the only data-driven criterion for choosing $\lambda$ is cross-validation. In this setting, the graph structure of the data makes cross-validation computationally too costly since one would need to repeat the cross-validation over a large set of test/train sets.
Consequently, we use model selection criteria. We consider the BIC \citet{Schwarz1978EstimatingDimensionModel}, the AIC \citet{akaike1974NewLookStatistical}, and generalized cross-validation (GCV) \citep[Section 7]{Hastie2009ElementsStatisticalLearning}:
\begin{align*}
    \text{BIC}(\lambda) &= 2 \ell(\hat{\bm{\theta}}) + \log(p) e(\lambda),\\
    \text{AIC}(\lambda) &= 2 \ell(\hat{\bm{\theta}}) + 2 e(\lambda),\\
    \text{GCV}(\lambda) &= \frac{2\ell(\hat{\bm{\theta}})}{p(1 - e(\lambda) / p) ^2},
\end{align*}
where $e(\lambda)$ is the effective model dimension.

The effective dimension is a generalization of the number of parameters to linear fitting methods: if the estimate writes $\hat{\bm{\theta}} = \mathbf{S}(\lambda) \bm{x}$ with $\mathbf{S}(\lambda)$ independent from $\bm{x}$, then we define $e(\lambda) = \textbf{Tr}(\mathbf{S}(\lambda))$. When $\mathbf{S}$ is a projection (as in the unpenalized linear model), the effective dimension is the number of parameters in the model. In the general case, this formula accounts for both the number of selected parameters and the shrinkage between the parameters. Besides allowing to use cost-effective selection criteria, $e(\lambda)$ is important in itself: it allows to quantify the degree of freedom of the estimate.

We now derive the effective dimension for the adaptive ridge. As explained in \citet[Section~S1]{goepp2021RegularizedBidimensionalEstimation}, the adaptive ridge Equations \eqref{eq:ar_algo1} and \eqref{eq:ar_algo2} is a numerical scheme which minimizes the penalized problem
\begin{equation*}
  \min_{\bm{\theta}}\phantom{,} \ell(\bm{\theta}) + \frac{\lambda}{2} \sum_{j \sim k} \log((\theta_j - \theta_k) ^ 2 + \varepsilon)
\end{equation*}
through \citet{Fan2001VariableSelectionNonconcave}'s one-step approximation procedure called local quadratic approximation (LQA).
Now following the work of \citet{Fan2001VariableSelectionNonconcave, Fan2006StatisticalChallengesHigh} for extending the notion of effective dimension to this class of estimating algorithms, we use
\begin{equation}
    e(\lambda) \triangleq \text{Tr}\left((\mathbf{\Sigma}^{-1} + \lambda \bm{K}^{(l)})^{-1}\mathbf{\Sigma}^{-1}\right)
    \label{eq:effective_num_param}
\end{equation}
when the iteration step $(l)$ is at convergence.
There is no theoretical justification for this formula, which is rather is motivated by the analogy with linear smoothers, since our estimate takes the form $\bm\theta = (\bm\Sigma^{-1} + \lambda \bm K^{(l)})^{-1}\bm{\Sigma}^{-1}\bm x$, when $l$ is large enough that the method has converged.

Computing $e(\lambda)$ only requires solving the linear system in the right-hand side of \eqref{eq:effective_num_param}, which comes at a small computational cost since the Cholesky decomposition of $\mathbf{\Sigma}^{-1} + \lambda \bm{K}^{(l)}$ is already in memory at convergence of the algorithm.

\subsection{Application}


\subsubsection{Simulation study}

\begin{figure}
\centering
\subfloat[Dataset 1: $p=390$, $q = 12$]{
\includegraphics[width = 0.45\textwidth]{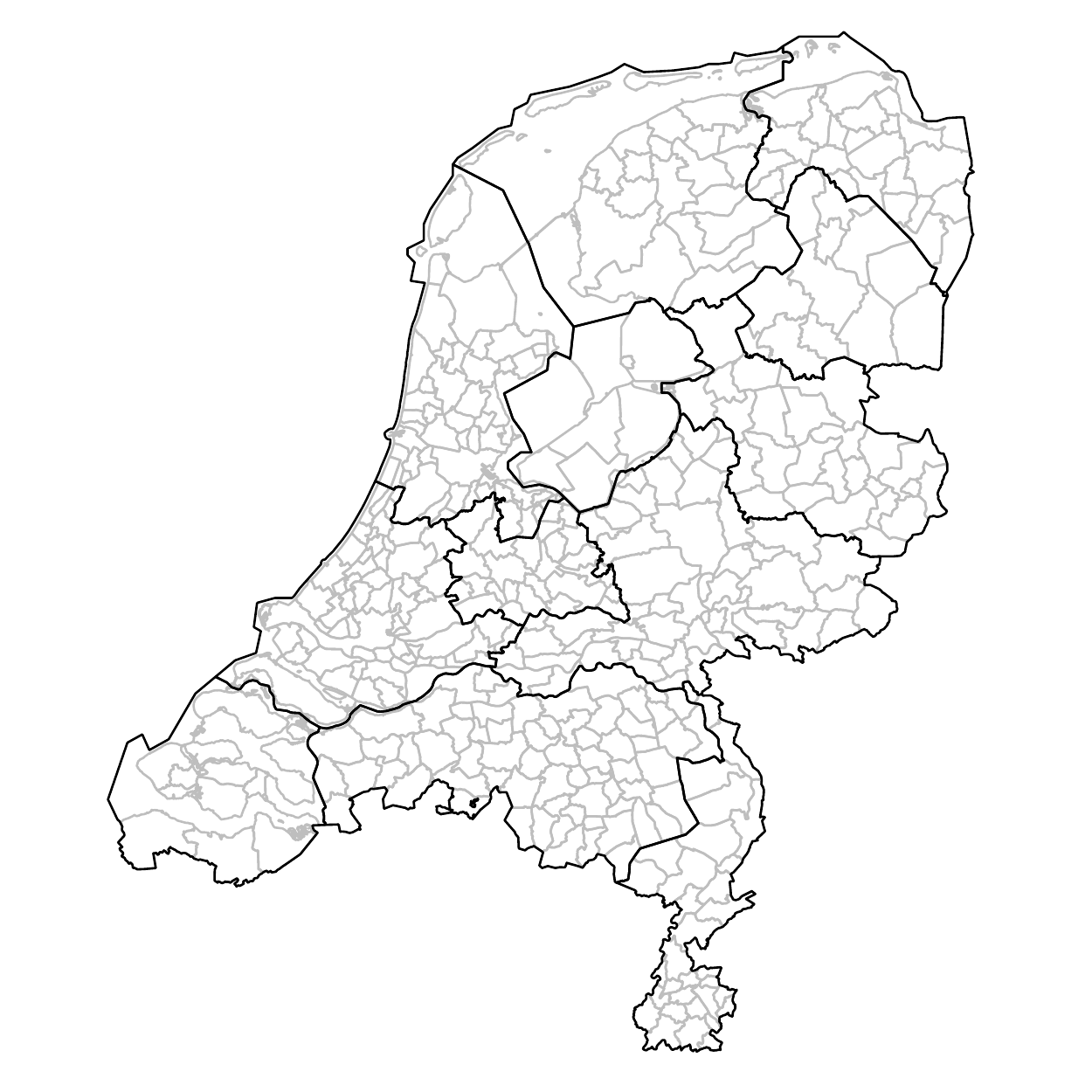}
\label{fig:dataset1_zones}
}
\hfil
\subfloat[Dataset 2: $p = 650$, $q = 99$]{
\includegraphics[width = 0.45\textwidth]{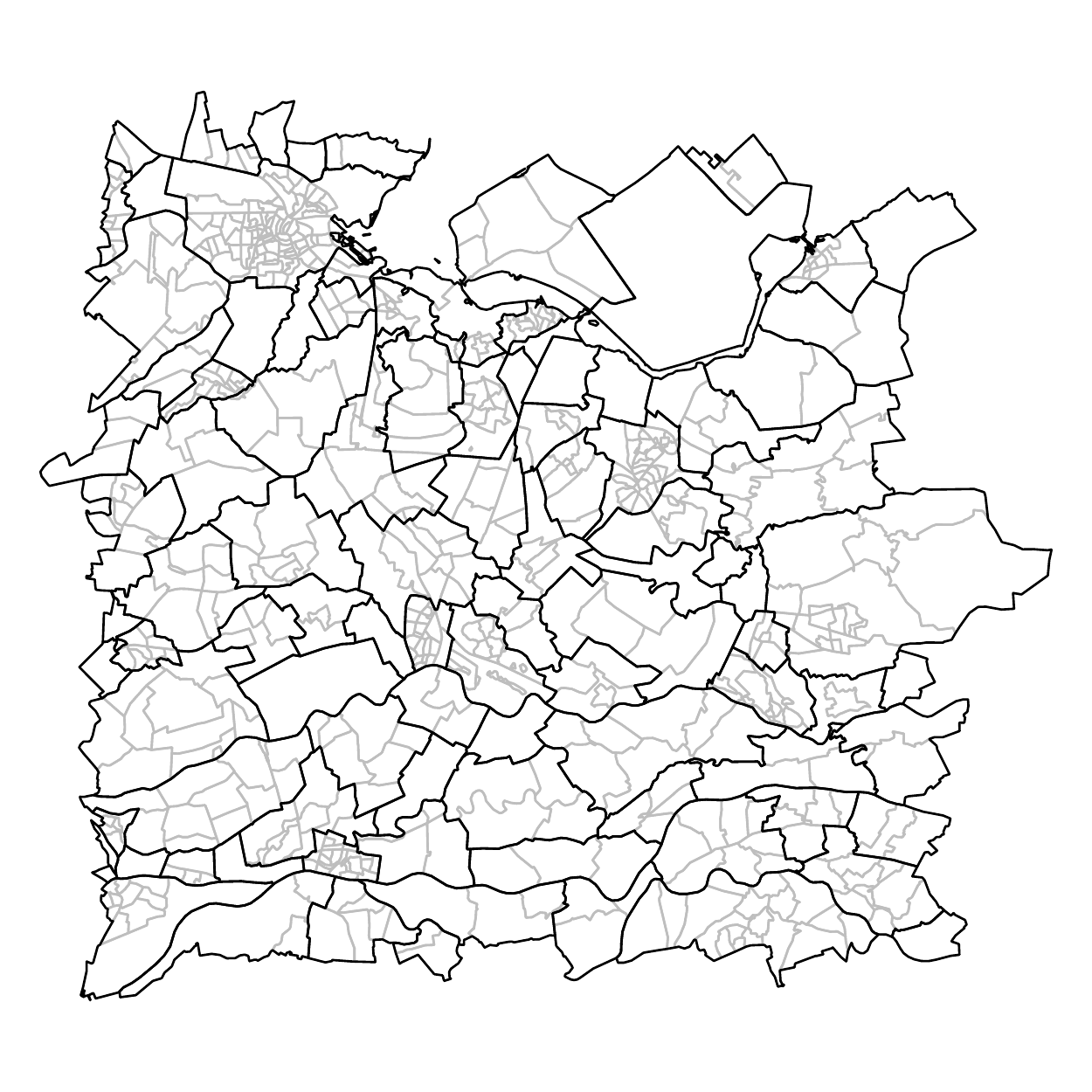}
\label{fig:dataset2_zones}
}
\hfil
\subfloat[Dataset 3: $p = 2955$, $q = 6$]{
\includegraphics[width = 0.45\textwidth]{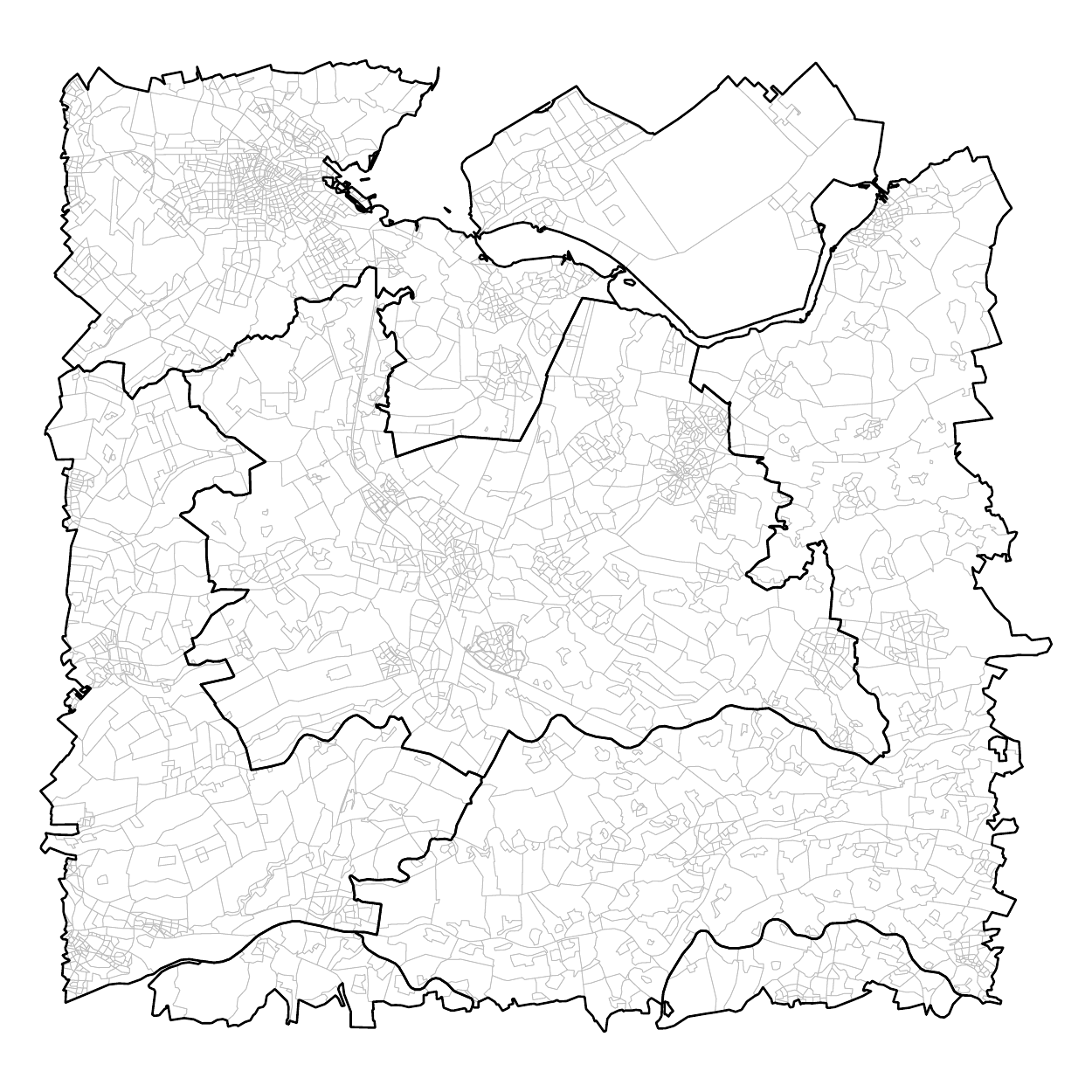}
\label{fig:dataset3_zones}
}
\hfil
\subfloat[Dataset 4: $p = 2955$, $q = 120$]{
\includegraphics[width = 0.45\textwidth]{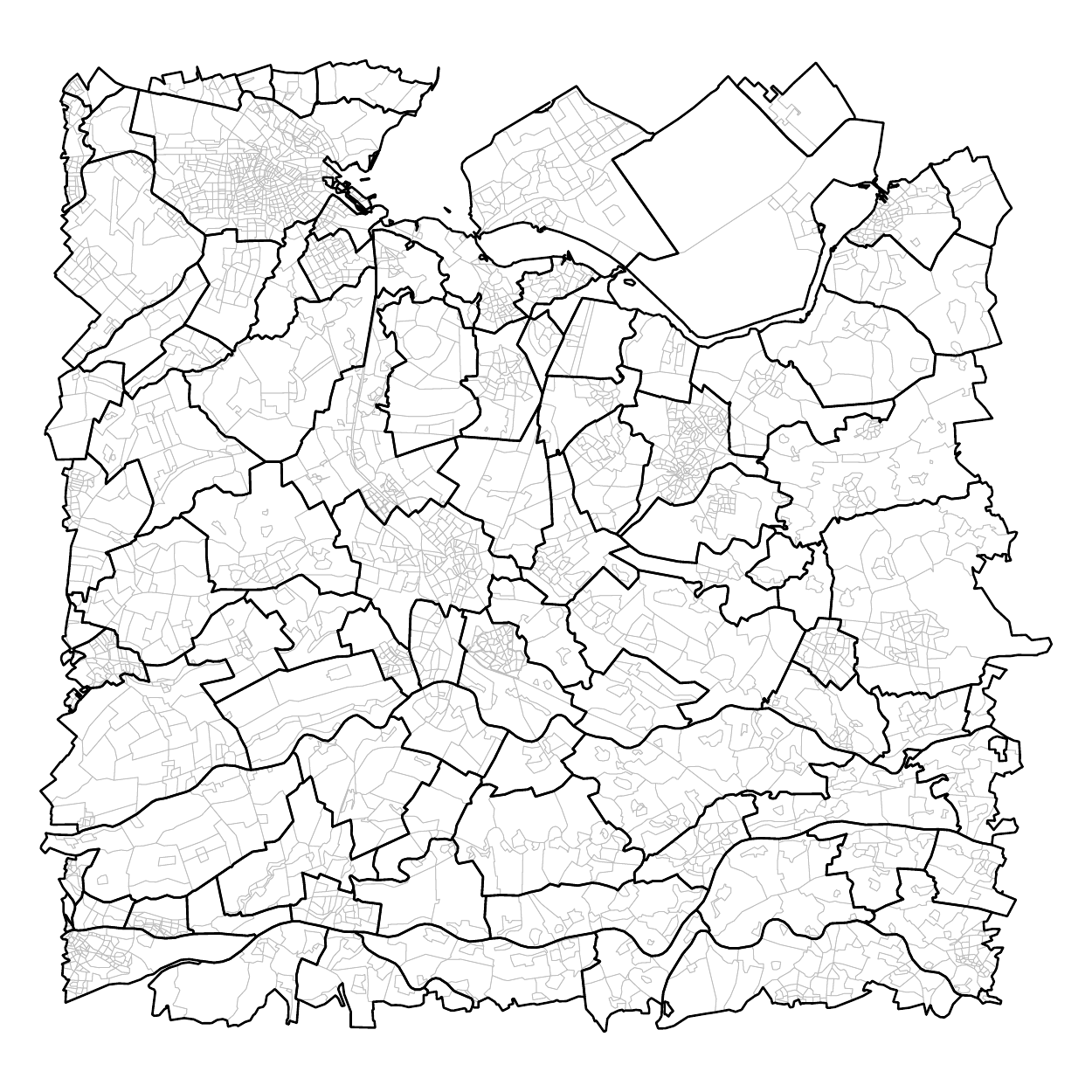}
\label{fig:dataset4_zones}
}
\hfil
\subfloat[Dataset 5: $p = 2955$, $q = 650$]{
\includegraphics[width = 0.45\textwidth]{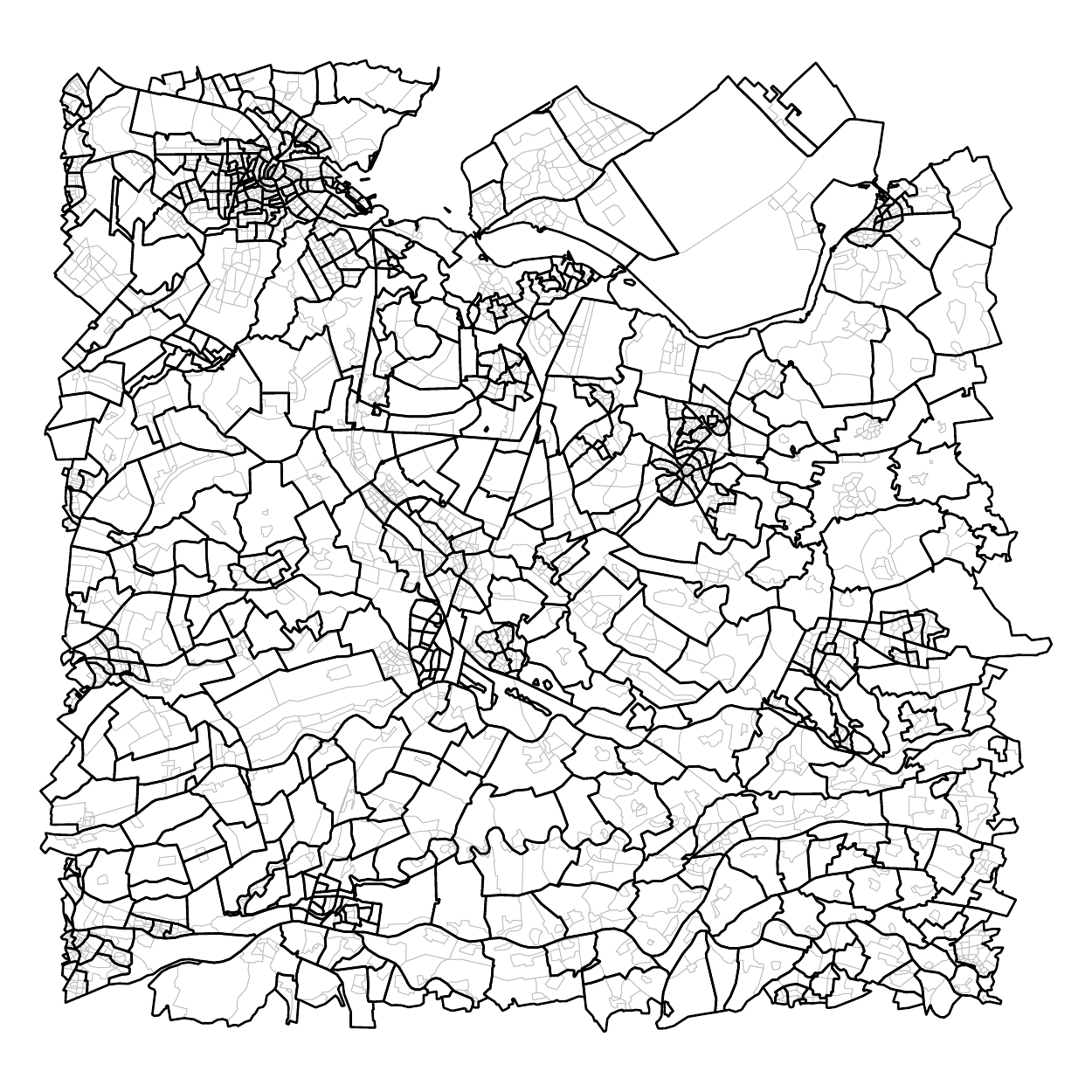}
\label{fig:dataset5_zones}
}
\hfil
\subfloat[Dataset 6: $p = 12920$, $q = 390$]{
\includegraphics[width = 0.45\textwidth]{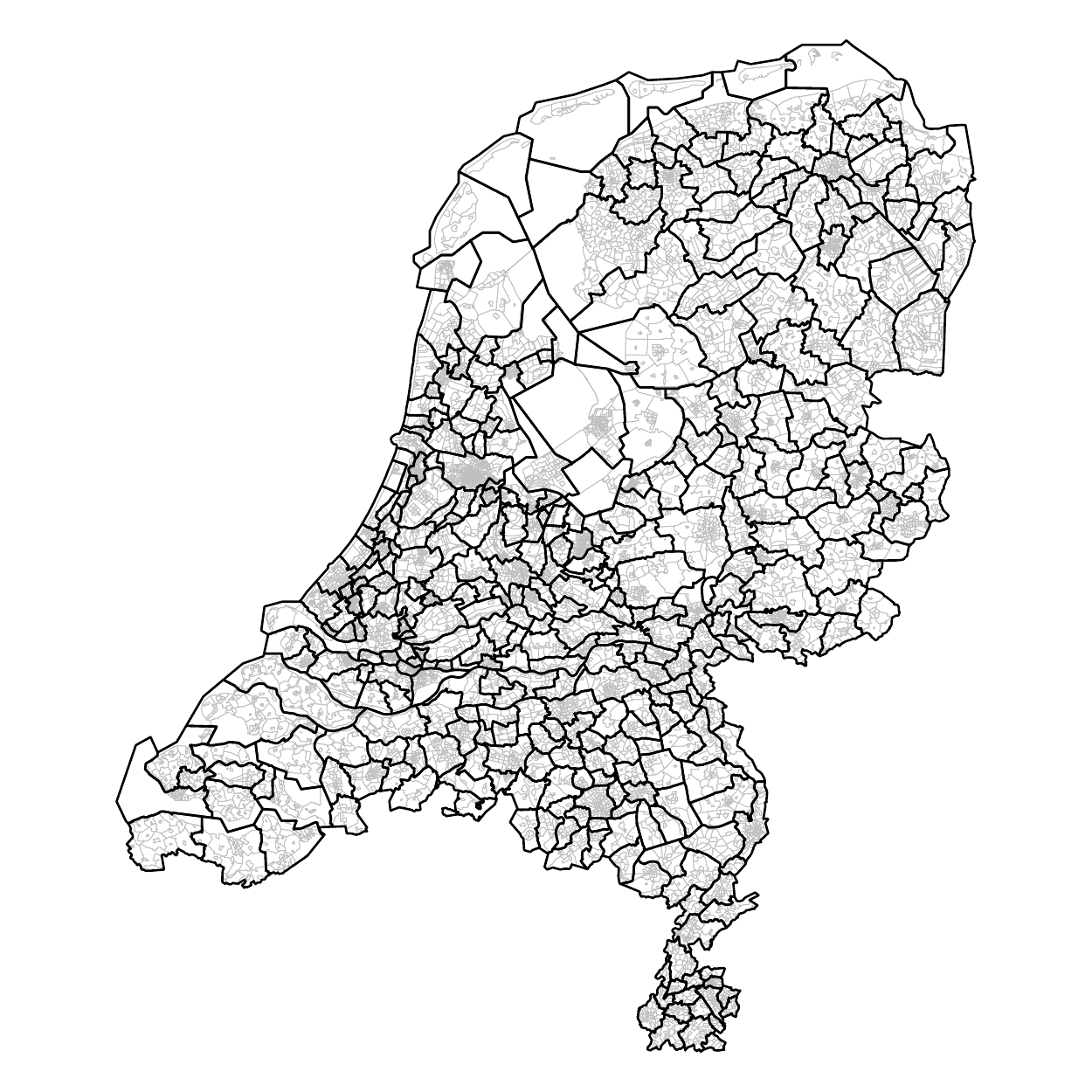}
\label{fig:dataset6_zones}
}
\caption{Areas (grey) and zones (black) of the 6 datasets used in simulation, with their numbers $p$ and $q$ respectively.}
\label{fig:dataset_zones}
\end{figure}

We illustrate our method in six simulation settings.
The graph structure used to define the zones and the adjacency between regions is taken from real data.
We use the spatial polygons defining several administrative divisions of the Netherlands: regions, municipalities, districts, and neighborhoods.
We consider different simulation settings, with varried numbers of areas ($p$) and zones ($q$).
We select 6 settings, with $p$ varrying from 390 to 12920 and $q$ varrying from $6$ to $390$.
The datasets are represented in Figure~\ref{fig:dataset_zones} and their properties are summarized in Table~\ref{tab:dataset_info}.
We report the average number of areas per zone, which indicates how hard the estimation of the areas and their spatial effect is: the more areas in a zone, the more information there is about its underlying true value.
This dataset based on real data is further detailed in Section~\ref{section:methods_real_data} and Figure~\ref{fig:reference_map}.

The data $\bm{x}$ is simulated from Eq.~\eqref{eq:model_gaussian}, where the true parameter $\bm \theta$ is constant over each zone and the errors are homoskedastic ($\bm\Sigma = \sigma \bm{I}$). The true parameters $\bm\theta$ are set equal on each zone and are generated as iid Poisson samples of parameter $10$. The noise variance $\sigma^2$ is set to a series of values between $0.1$ and $5$.

We select the best $\lambda$ over a grid of 50 values using either of the three criteria. 
We run the graph-fused lasso using the implementation from the R package \texttt{flsa} \citep{hoefling2020FlsaPathAlgorithm}.
We run flsa on the same sequence of values of~$\lambda$.

\begin{table}

\caption{\label{tab:dataset_info}Information on the different simulation designs.}
\centering
\begin{tabular}[t]{lrrr}
\toprule
\makecell[l]{Dataset \\ name} & \makecell[c]{Number of \\areas ($p$)} & \makecell[c]{Number of \\zones ($q$)}& \makecell[l]{Average number of \\ areas per zones}\\
\midrule
\cellcolor{gray!6}{Dataset 1} & \cellcolor{gray!6}{390} & \cellcolor{gray!6}{12} & \cellcolor{gray!6}{32.50}\\
Dataset 2 & 650 & 99 & 6.57\\
\cellcolor{gray!6}{Dataset 3} & \cellcolor{gray!6}{2955} & \cellcolor{gray!6}{6} & \cellcolor{gray!6}{492.00}\\
Dataset 4 & 2955 & 99 & 29.80\\
\cellcolor{gray!6}{Dataset 5} & \cellcolor{gray!6}{2955} & \cellcolor{gray!6}{650} & \cellcolor{gray!6}{4.55}\\
Dataset 6 & 12920 & 390 & 33.13\\
\bottomrule
\end{tabular}
\end{table}

\subsubsection{Real data: overweight in the Netherlands}\label{section:methods_real_data}

\begin{figure}
    \centering
    \includegraphics[width = \linewidth, keepaspectratio]{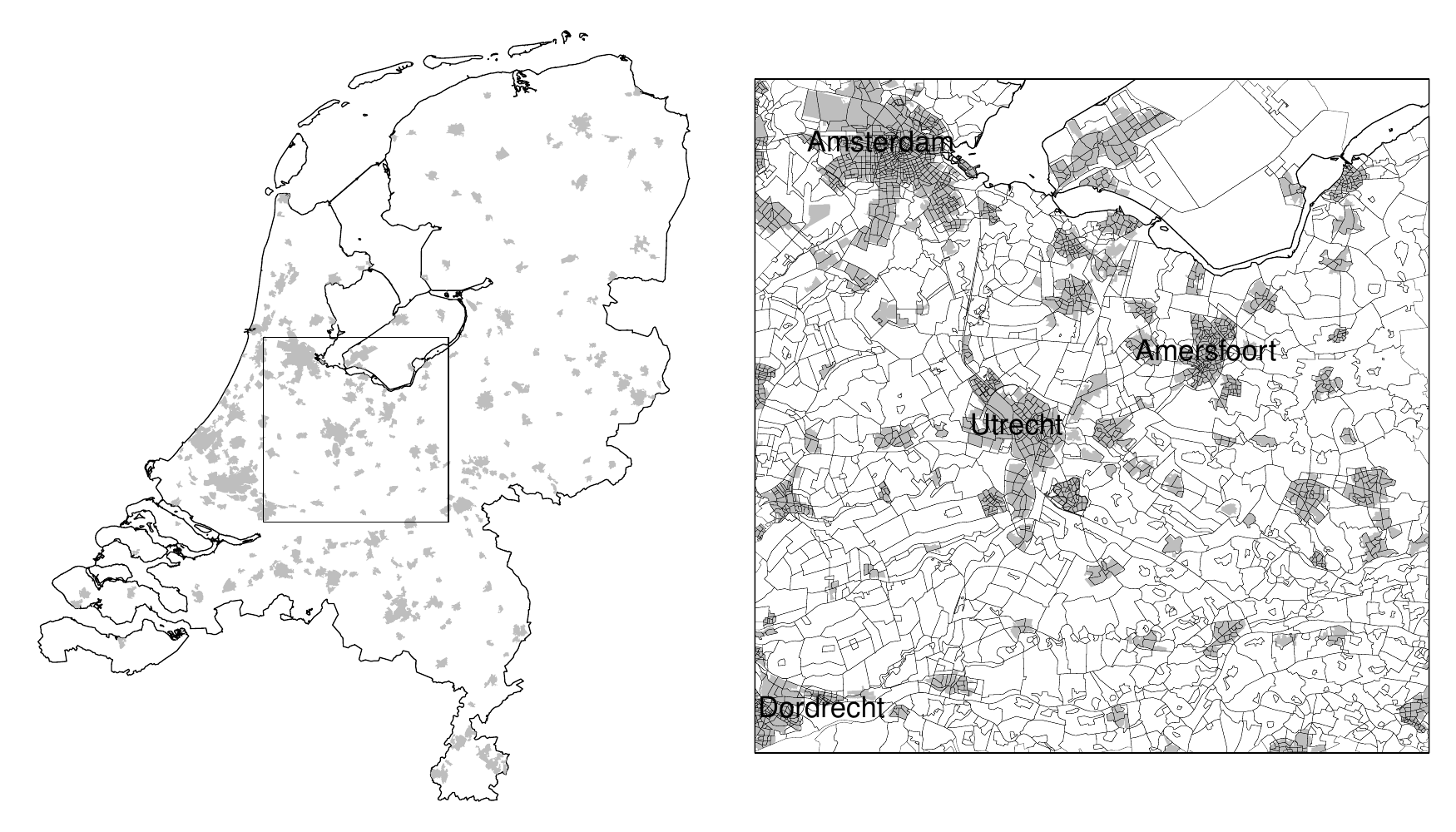}
    \caption{The square~$75$ by~$75$ kilometers region of interest located near the centre of the Netherlands. For illustration and orientation, the grey shaded areas are large populated places, e.g. cities like Amsterdam and Utrecht. The map on the right shows the 2,955 neighborhoods in detail.}
    \label{fig:reference_map}
\end{figure}

For our real-data application we focus on the estimated overweight prevalence at neighborhood level in the Netherlands in 2016.
Overweight is being defined as having a body mass index between 20 and 25.
We consider a square region of~$75$ by~$75$ kilometers near the center of the Netherlands (the same that is used in simulation Datasets (f)-(e), c.f. Figure~\ref{fig:dataset_zones}).
Figure~\ref{fig:reference_map} shows the region on a map.
The region consists of 2,955 neighborhoods.
In the Netherlands, neighborhoods are defined for administrative use by municipalities and data collection by Statistics Netherlands (CBS).
Neighbourhoods are coherent regions that are based on several characteristics like age, geographical barriers such as busy roads, having similar urban and/or architectural features, or having similar functional, social or political characteristics.
Neighbourhoods have no formal status.

\begin{figure}
    \centering
    \includegraphics[width = \linewidth, keepaspectratio]{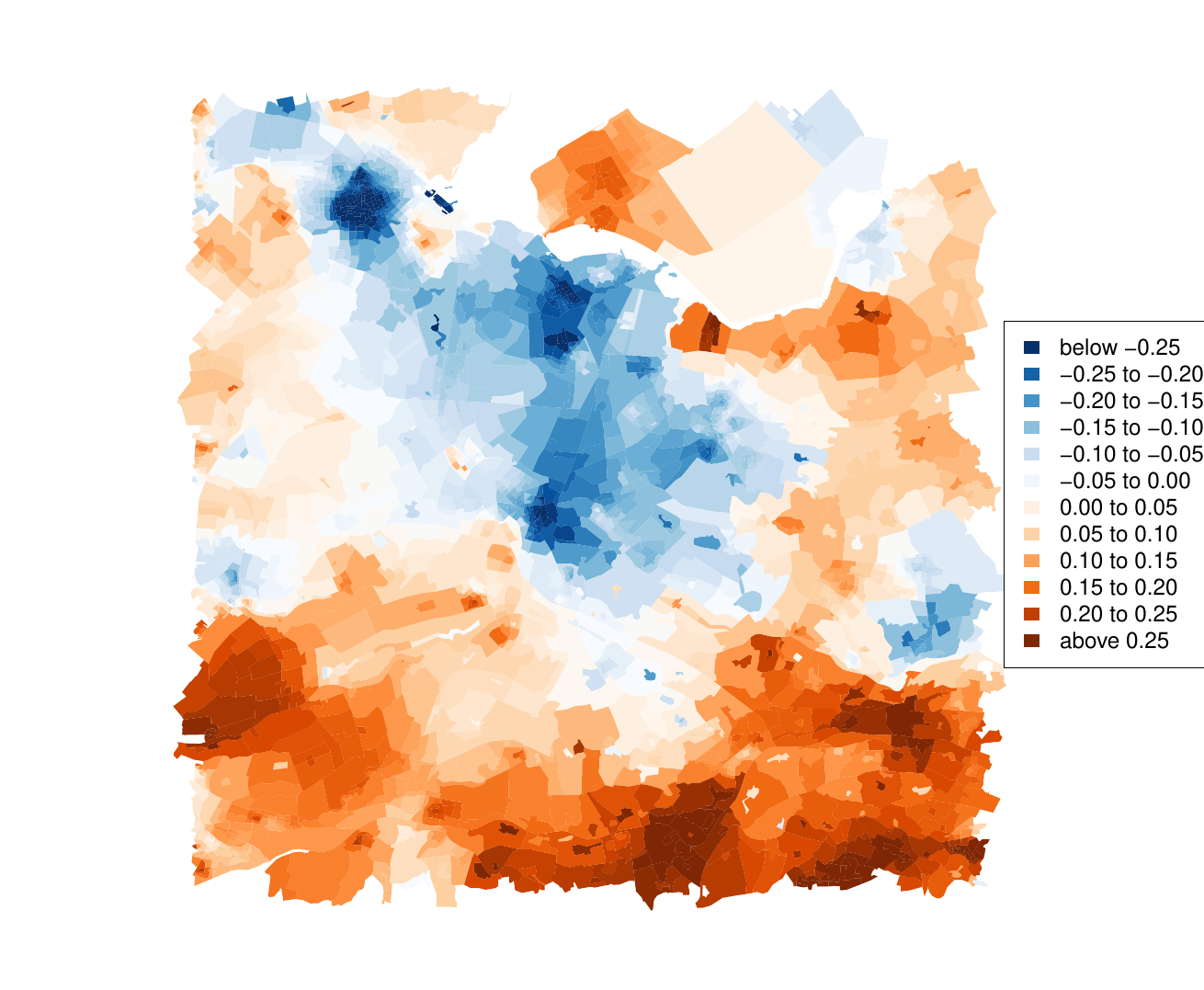}
    \caption{Unsegmented spatial effect for overweight for the 2,955 neighborhoods as estimated by the small area estimation model. Blue colours indicate lower log-odds compared to the expected log-odds, orange colours higher log-odds.}
    \label{fig:utrecht_mrf_unseg}
\end{figure}

The goal is to identify spatial zones consisting of adjacent neighborhoods that have higher or lower prevalence than can be expected based on the demographic and socio-economic characteristics of neighborhoods alone.

We follow a two-step procedure. First, we fit the small area estimation model by \citep{vandeKassteele2017Estimatingprevalence26} to our sample data and extract the estimated spatial effect of each neighborhood, given as log-odds ratios, as well as the covariance matrix. Next, we perform the spatial segmentation as described above.

Figure \ref{fig:utrecht_mrf_unseg} shows the unsegmented spatial random effect for the 2,955 neighborhoods in terms of log-odds ratio's. Blue colours indicate lower than expected log-odds on overweight. Orange colours indicate higher than expected log-odds on overweight. Already clear patterns can be seen, e.g. in large cities like Amsterdam, Utrecht and Amersfoort, the overweight prevalence is lower than expected based on demographic and socio-economic characteristics. In the rural areas in the south, the prevalence is higher than expected.

We used an updated version of the small area estimation model as described by van de Kassteele et al (2017) \citep{vandeKassteele2017Estimatingprevalence26}. The updated model has several improvements. First, sample data are from 2016 instead of 2012. Second, the model includes educational level as a predictor variable. Third, more two-way interactions are included: age by sex, age by ethnicity, age by marital status, age by educational level, sex by ethnicity, sex by marital status and sex by educational level. Fourth, all predictor variables, both numeric as categorical, entered the model using basis functions and penalization of the regression coefficients. This also enabled automated feature selection, i.e. predictors that are not relevant will not be selected in the model, resulting in a more parsimonious model.

Our method requires the precision matrix as input. We have used the following two-step process to estimate this matrix. First, the covariance matrix $\bm\Sigma$ of the spatial effect term was extracted from the small area estimation model. Next, we used the graphical lasso \citep{Friedman2008SparseInverseCovariance}, as implemented in the package \texttt{huge} \citep{Zhao2012hugePackageHighdimensional}, to generate a sparse estimate of $\bm\Sigma^{-1}$. 

\section{Results}

\subsection{Simulation study}\label{section:results_simulation}
\paragraph{The adaptive ridge produces sparser estimates}
\begin{figure}
\centering
\subfloat[RMSE as a function of $\lambda$]{
\includegraphics[width = 0.46\textwidth]{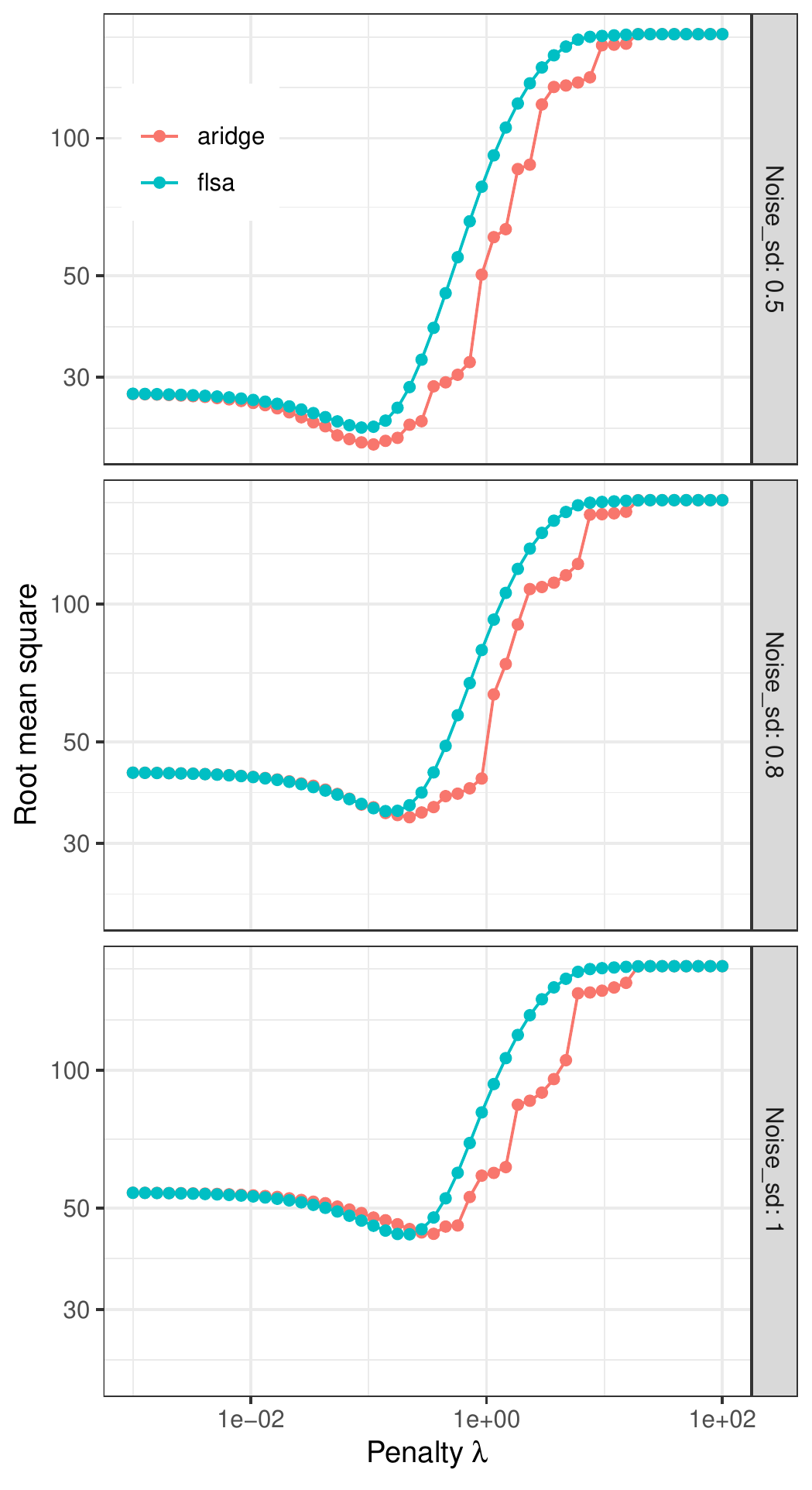}
\label{fig:utrecht_neigh_pc_district_rms_lambda}
}
\hfil
\subfloat[RMSE as a function of the model dimension]{
\includegraphics[width = 0.46\textwidth]{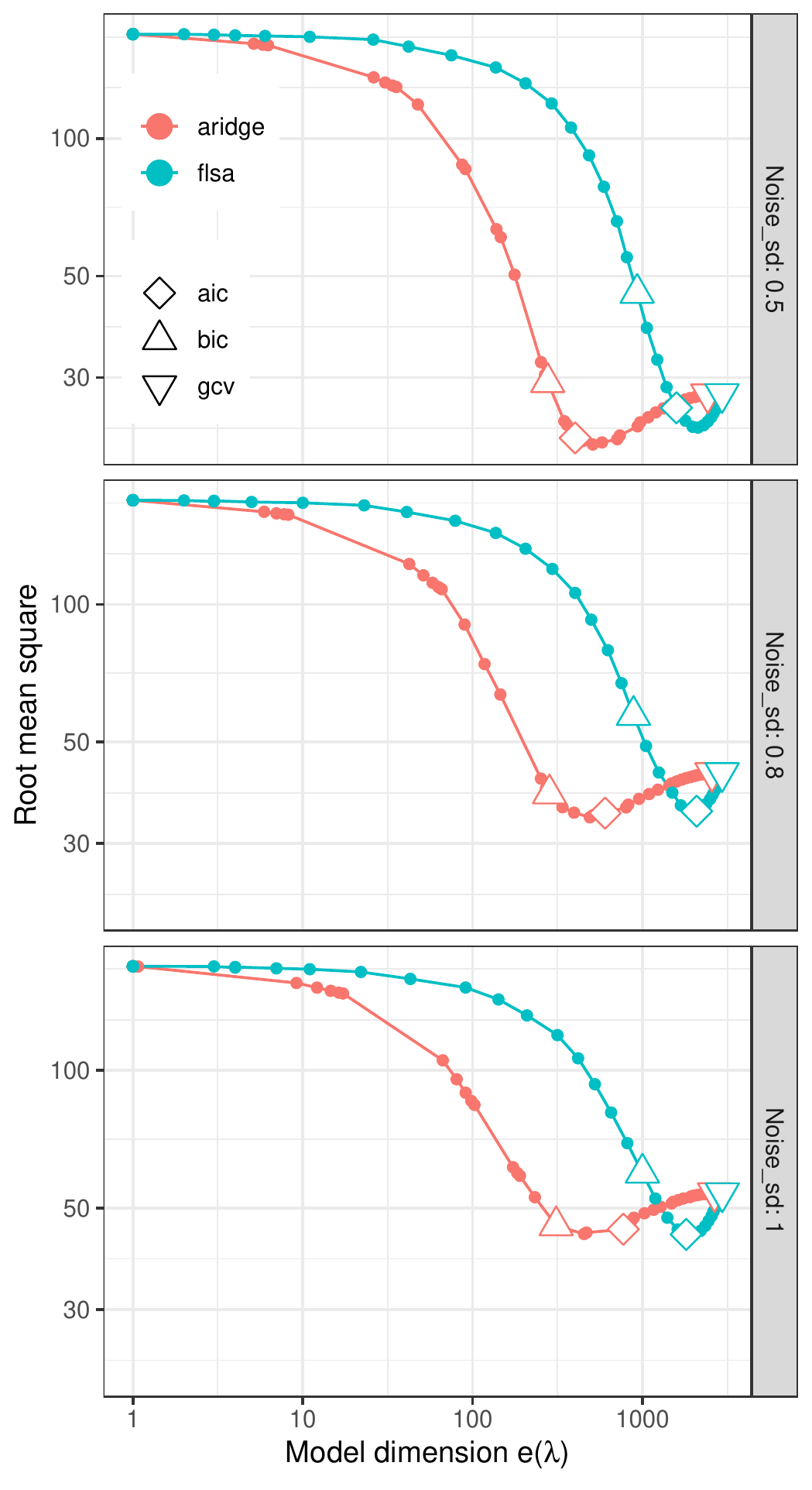}
\label{fig:utrecht_neigh_pc_district_rms_model_dim_crit}
}
\caption{Root mean squared error (RMSE) of one estimate in Dataset 5 as a function of (a) the penalty~$\lambda$ and (b) the model dimension~$e(\lambda)$. The left-hand figure shows that at the optimal $\lambda$, the adaptive ridge yields an estimate with a RMSE close or slightly less than that of flsa. The right-hand figure shows that it yields a way sparser model (for the same amount of RMSE, since both curves have similar minimal values). For both methods, the AIC criterion is best at selecting the optimal $\lambda$.}
\label{fig:utrecht_neigh_pc_district_rms}
\end{figure}

We compare the adaptive ridge and the flsa on a single estimate of Dataset 5 ($p = 2955$, $q = 650$) in Figure~\ref{fig:utrecht_neigh_pc_district_rms}, for $\sigma = 0.5$, $0.8$, and $1$. We present the root mean squared error (RMSE) for varying values of $\lambda$. Figure~\ref{fig:utrecht_neigh_pc_district_rms}\emph{a} shows that the adaptive ridge performs always better than, or as well as, flsa for the same amount of regularization and Figure~\ref{fig:utrecht_neigh_pc_district_rms}\emph{b} shows that \emph{i)} for equal RMSE, the adaptive ridge selects sparser models and \emph{ii)} the AIC criterion is best at selecting the optimal $\lambda$. Similar results were observed with the other simulation designs (results not shown here). Extensive simulations (not shown here) across a larger array of values of $\sigma$ shows that for small $\sigma$, GCV outperforms AIC and BIC and for large $\sigma$, AIC and BIC are close and outperform GCV. In applications, we recommend using the AIC if estimation performance is prioritized and the BIC if model sparsity is prioritized.

\paragraph{The adaptive ridge yields better estimates in terms of RMSE}

Table~\ref{tab:rms_dim_table} compares adaptive ridge and flsa estimates in terms of RMSE and model dimension across all values of $\sigma$ and all simulation settings. The adaptive ridge has better estimation performance than flsa (lower RMSE) when $\sigma$ is not too high ($\leq 1.1$), in which case it has slightly worse but comparable performance. Moreover, it consistently fits sparser estimates.

\begingroup\fontsize{7}{9}\selectfont

\begin{longtable}[t]{llllll}
\caption{\label{tab:rms_dim_table}Performance estimation and selected number of zones of the adaptive ridge 
        compared to Hoefling's fused lasso signal approximator (lfsa), using the AIC.}\\
\toprule
\multicolumn{1}{c}{ } & \multicolumn{2}{c}{\makecell[c]{Model \\ dimension}} & \multicolumn{3}{c}{RMSE} \\
\cmidrule(l{3pt}r{3pt}){2-3} \cmidrule(l{3pt}r{3pt}){4-6}
\makecell[c]{Noise \\ standard \\ deviation} & \makecell[c]{Adaptive \\ Ridge} & FLSA & \makecell[c]{Adaptive \\ Ridge} & FLSA & \makecell[c]{No \\ Regula- \\ rization}\\
\midrule
\endfirsthead
\caption[]{Performance estimation and selected number of zones of the adaptive rid \textit{(continued)}}\\
\toprule
\multicolumn{1}{c}{ } & \multicolumn{2}{c}{\makecell[c]{Model \\ dimension}} & \multicolumn{3}{c}{RMSE} \\
\cmidrule(l{3pt}r{3pt}){2-3} \cmidrule(l{3pt}r{3pt}){4-6}
\makecell[c]{Noise \\ standard \\ deviation} & \makecell[c]{Adaptive \\ Ridge} & FLSA & \makecell[c]{Adaptive \\ Ridge} & FLSA & \makecell[c]{No \\ Regula- \\ rization}\\
\midrule
\endhead

\endfoot
\bottomrule
\endlastfoot
\addlinespace[0.3em]
\multicolumn{6}{l}{\textbf{Dataset 1}}\\
\hspace{1em}\cellcolor{gray!6}{0.1} & \cellcolor{gray!6}{\boldmath{$ 11.1 \pm 0.0214 $}} & \cellcolor{gray!6}{$ 43 \pm 1.41 $} & \cellcolor{gray!6}{\boldmath{$ 0.0815 \pm 0.00297 $}} & \cellcolor{gray!6}{$ 0.313 \pm 0.0685 $} & \cellcolor{gray!6}{$ 0 $}\\
\hspace{1em}0.3 & \boldmath{$ 11.4 \pm 0.236 $} & $ 41.5 \pm 2.12 $ & \boldmath{$ 0.123 \pm 0.00747 $} & $ 0.408 \pm 0.0858 $ & $ 0 $\\
\hspace{1em}\cellcolor{gray!6}{0.5} & \cellcolor{gray!6}{\boldmath{$ 15 \pm 3.01 $}} & \cellcolor{gray!6}{$ 36.5 \pm 4.95 $} & \cellcolor{gray!6}{\boldmath{$ 0.253 \pm 0.0264 $}} & \cellcolor{gray!6}{$ 0.505 \pm 0.0236 $} & \cellcolor{gray!6}{$ 0 $}\\
\hspace{1em}0.7 & \boldmath{$ 16.2 \pm 3.58 $} & $ 40.5 \pm 0.707 $ & \boldmath{$ 0.358 \pm 0.0971 $} & $ 0.433 \pm 0.0979 $ & $ 0 $\\
\hspace{1em}\cellcolor{gray!6}{0.9} & \cellcolor{gray!6}{\boldmath{$ 16.7 \pm 4.12 $}} & \cellcolor{gray!6}{$ 48.5 \pm 4.95 $} & \cellcolor{gray!6}{\boldmath{$ 0.467 \pm 0.0335 $}} & \cellcolor{gray!6}{$ 0.5 \pm 0.00638 $} & \cellcolor{gray!6}{$ 0 $}\\
\hspace{1em}1.1 & \boldmath{$ 26.4 \pm 13 $} & $ 49.5 \pm 4.95 $ & $ 0.699 \pm 0.0717 $ & \boldmath{$ 0.614 \pm 0.0112 $} & $ 0 $\\
\hspace{1em}\cellcolor{gray!6}{2} & \cellcolor{gray!6}{$ 75.4 \pm 0.109 $} & \cellcolor{gray!6}{\boldmath{$ 73.5 \pm 31.8 $}} & \cellcolor{gray!6}{$ 1.61 \pm 0.0144 $} & \cellcolor{gray!6}{\boldmath{$ 0.922 \pm 0.0976 $}} & \cellcolor{gray!6}{$ 0 $}\\
\hspace{1em}5 & \boldmath{$ 153 \pm 13.6 $} & $ 362 \pm 2.12 $ & \boldmath{$ 4.6 \pm 0.0833 $} & $ 4.67 \pm 0.0421 $ & $ 0 $\\
\addlinespace[0.3em]
\multicolumn{6}{l}{\textbf{Dataset 2}}\\
\hspace{1em}\cellcolor{gray!6}{0.1} & \cellcolor{gray!6}{\boldmath{$ 45.3 \pm 1.21 $}} & \cellcolor{gray!6}{$ 146 \pm 0.707 $} & \cellcolor{gray!6}{$ 0.356 \pm 0.0186 $} & \cellcolor{gray!6}{\boldmath{$ 0.322 \pm 0.00223 $}} & \cellcolor{gray!6}{$ 2.5 $}\\
\hspace{1em}0.3 & \boldmath{$ 44.8 \pm 0.549 $} & $ 162 \pm 4.95 $ & \boldmath{$ 0.392 \pm 0.00749 $} & $ 0.525 \pm 0.0935 $ & $ 7.6 $\\
\hspace{1em}\cellcolor{gray!6}{0.5} & \cellcolor{gray!6}{\boldmath{$ 49.1 \pm 4.04 $}} & \cellcolor{gray!6}{$ 146 \pm 4.24 $} & \cellcolor{gray!6}{\boldmath{$ 0.402 \pm 0.0213 $}} & \cellcolor{gray!6}{$ 0.794 \pm 0.0157 $} & \cellcolor{gray!6}{$ 13 $}\\
\hspace{1em}0.7 & \boldmath{$ 38.5 \pm 0.0574 $} & $ 150 \pm 6.36 $ & \boldmath{$ 0.613 \pm 0.0326 $} & $ 0.809 \pm 0.0163 $ & $ 18 $\\
\hspace{1em}\cellcolor{gray!6}{0.9} & \cellcolor{gray!6}{\boldmath{$ 50.7 \pm 2.07 $}} & \cellcolor{gray!6}{$ 156 \pm 19.1 $} & \cellcolor{gray!6}{\boldmath{$ 0.683 \pm 0.0774 $}} & \cellcolor{gray!6}{$ 0.946 \pm 0.19 $} & \cellcolor{gray!6}{$ 24 $}\\
\hspace{1em}1.1 & \boldmath{$ 68.9 \pm 0.522 $} & $ 157 \pm 26.9 $ & \boldmath{$ 0.771 \pm 0.0132 $} & $ 0.956 \pm 0.142 $ & $ 28 $\\
\hspace{1em}\cellcolor{gray!6}{2} & \cellcolor{gray!6}{\boldmath{$ 128 \pm 26.8 $}} & \cellcolor{gray!6}{$ 332 \pm 55.2 $} & \cellcolor{gray!6}{$ 1.6 \pm 0.0531 $} & \cellcolor{gray!6}{\boldmath{$ 1.31 \pm 0.0421 $}} & \cellcolor{gray!6}{$ 51 $}\\
\hspace{1em}5 & \boldmath{$ 260 \pm 33 $} & $ 599 \pm 9.9 $ & $ 4.66 \pm 0.277 $ & \boldmath{$ 4.64 \pm 0.325 $} & $ 130 $\\
\addlinespace[0.3em]
\multicolumn{6}{l}{\textbf{Dataset 3}}\\
\hspace{1em}\cellcolor{gray!6}{0.1} & \cellcolor{gray!6}{\boldmath{$ 60.6 \pm 0.135 $}} & \cellcolor{gray!6}{$ 298 \pm 2.83 $} & \cellcolor{gray!6}{\boldmath{$ 0.178 \pm 0.00229 $}} & \cellcolor{gray!6}{$ 0.213 \pm 0.000189 $} & \cellcolor{gray!6}{$ 5.5 $}\\
\hspace{1em}0.3 & \boldmath{$ 61.6 \pm 1.56 $} & $ 282 \pm 12 $ & \boldmath{$ 0.185 \pm 0.0031 $} & $ 0.404 \pm 0.00924 $ & $ 16 $\\
\hspace{1em}\cellcolor{gray!6}{0.5} & \cellcolor{gray!6}{\boldmath{$ 61.3 \pm 3.64 $}} & \cellcolor{gray!6}{$ 286 \pm 2.12 $} & \cellcolor{gray!6}{\boldmath{$ 0.241 \pm 0.0106 $}} & \cellcolor{gray!6}{$ 0.544 \pm 0.00403 $} & \cellcolor{gray!6}{$ 28 $}\\
\hspace{1em}0.7 & \boldmath{$ 67 \pm 4.07 $} & $ 306 \pm 0.707 $ & \boldmath{$ 0.331 \pm 0.00976 $} & $ 0.568 \pm 0.0106 $ & $ 38 $\\
\hspace{1em}\cellcolor{gray!6}{0.9} & \cellcolor{gray!6}{\boldmath{$ 89 \pm 24.3 $}} & \cellcolor{gray!6}{$ 254 \pm 17.7 $} & \cellcolor{gray!6}{\boldmath{$ 0.41 \pm 0.00876 $}} & \cellcolor{gray!6}{$ 0.703 \pm 0.00372 $} & \cellcolor{gray!6}{$ 48 $}\\
\hspace{1em}1.1 & \boldmath{$ 141 \pm 58.2 $} & $ 291 \pm 4.24 $ & \boldmath{$ 0.57 \pm 0.0662 $} & $ 0.713 \pm 0.000775 $ & $ 59 $\\
\hspace{1em}\cellcolor{gray!6}{2} & \cellcolor{gray!6}{$ 543 \pm 65.8 $} & \cellcolor{gray!6}{\boldmath{$ 486 \pm 3.54 $}} & \cellcolor{gray!6}{$ 1.57 \pm 0.0246 $} & \cellcolor{gray!6}{\boldmath{$ 0.878 \pm 0.0277 $}} & \cellcolor{gray!6}{$ 110 $}\\
\hspace{1em}5 & \boldmath{$ 1010 \pm 143 $} & $ 2630 \pm 12 $ & $ 4.55 \pm 0.0702 $ & \boldmath{$ 4.49 \pm 0.019 $} & $ 270 $\\
\addlinespace[0.3em]
\multicolumn{6}{l}{\textbf{Dataset 4}}\\
\hspace{1em}\cellcolor{gray!6}{0.1} & \cellcolor{gray!6}{\boldmath{$ 61 \pm 0.947 $}} & \cellcolor{gray!6}{$ 298 \pm 2.83 $} & \cellcolor{gray!6}{\boldmath{$ 0.176 \pm 0.0082 $}} & \cellcolor{gray!6}{$ 0.213 \pm 0.000189 $} & \cellcolor{gray!6}{$ 5.5 $}\\
\hspace{1em}0.3 & \boldmath{$ 62.9 \pm 2.58 $} & $ 282 \pm 12 $ & \boldmath{$ 0.175 \pm 0.0194 $} & $ 0.404 \pm 0.00924 $ & $ 16 $\\
\hspace{1em}\cellcolor{gray!6}{0.5} & \cellcolor{gray!6}{\boldmath{$ 64.8 \pm 4.46 $}} & \cellcolor{gray!6}{$ 286 \pm 2.12 $} & \cellcolor{gray!6}{\boldmath{$ 0.237 \pm 0.0213 $}} & \cellcolor{gray!6}{$ 0.544 \pm 0.00403 $} & \cellcolor{gray!6}{$ 28 $}\\
\hspace{1em}0.7 & \boldmath{$ 70.4 \pm 9.32 $} & $ 306 \pm 0.707 $ & \boldmath{$ 0.325 \pm 0.0228 $} & $ 0.568 \pm 0.0106 $ & $ 38 $\\
\hspace{1em}\cellcolor{gray!6}{0.9} & \cellcolor{gray!6}{\boldmath{$ 90.7 \pm 19.6 $}} & \cellcolor{gray!6}{$ 254 \pm 17.7 $} & \cellcolor{gray!6}{\boldmath{$ 0.424 \pm 0.0215 $}} & \cellcolor{gray!6}{$ 0.703 \pm 0.00372 $} & \cellcolor{gray!6}{$ 48 $}\\
\hspace{1em}1.1 & \boldmath{$ 147 \pm 42.8 $} & $ 291 \pm 4.24 $ & \boldmath{$ 0.569 \pm 0.0412 $} & $ 0.713 \pm 0.000775 $ & $ 59 $\\
\hspace{1em}\cellcolor{gray!6}{2} & \cellcolor{gray!6}{$ 506 \pm 94.4 $} & \cellcolor{gray!6}{\boldmath{$ 486 \pm 3.54 $}} & \cellcolor{gray!6}{$ 1.53 \pm 0.0778 $} & \cellcolor{gray!6}{\boldmath{$ 0.878 \pm 0.0277 $}} & \cellcolor{gray!6}{$ 110 $}\\
\hspace{1em}5 & \boldmath{$ 1110 \pm 96.9 $} & $ 2630 \pm 12 $ & $ 4.58 \pm 0.0818 $ & \boldmath{$ 4.49 \pm 0.019 $} & $ 270 $\\
\addlinespace[0.3em]
\multicolumn{6}{l}{\textbf{Dataset 5}}\\
\hspace{1em}\cellcolor{gray!6}{0.1} & \cellcolor{gray!6}{\boldmath{$ 429 \pm 33.7 $}} & \cellcolor{gray!6}{$ 1130 \pm 14.1 $} & \cellcolor{gray!6}{\boldmath{$ 0.187 \pm 0.0637 $}} & \cellcolor{gray!6}{$ 0.216 \pm 0.00045 $} & \cellcolor{gray!6}{$ 5.5 $}\\
\hspace{1em}0.3 & \boldmath{$ 417 \pm 27.7 $} & $ 1240 \pm 10.6 $ & \boldmath{$ 0.272 \pm 0.0227 $} & $ 0.429 \pm 0.00182 $ & $ 16 $\\
\hspace{1em}\cellcolor{gray!6}{0.5} & \cellcolor{gray!6}{\boldmath{$ 508 \pm 14.6 $}} & \cellcolor{gray!6}{$ 1360 \pm 180 $} & \cellcolor{gray!6}{\boldmath{$ 0.391 \pm 0.00397 $}} & \cellcolor{gray!6}{$ 0.555 \pm 0.0629 $} & \cellcolor{gray!6}{$ 28 $}\\
\hspace{1em}0.7 & \boldmath{$ 610 \pm 41 $} & $ 1430 \pm 12 $ & \boldmath{$ 0.534 \pm 0.0165 $} & $ 0.652 \pm 0.00449 $ & $ 38 $\\
\hspace{1em}\cellcolor{gray!6}{0.9} & \cellcolor{gray!6}{\boldmath{$ 661 \pm 20.7 $}} & \cellcolor{gray!6}{$ 1710 \pm 247 $} & \cellcolor{gray!6}{\boldmath{$ 0.719 \pm 0.00217 $}} & \cellcolor{gray!6}{$ 0.748 \pm 0.0141 $} & \cellcolor{gray!6}{$ 48 $}\\
\hspace{1em}1.1 & \boldmath{$ 788 \pm 197 $} & $ 2000 \pm 28.3 $ & $ 0.921 \pm 0.0331 $ & \boldmath{$ 0.871 \pm 0.0218 $} & $ 59 $\\
\hspace{1em}\cellcolor{gray!6}{2} & \cellcolor{gray!6}{\boldmath{$ 1080 \pm 2.49 $}} & \cellcolor{gray!6}{$ 2560 \pm 103 $} & \cellcolor{gray!6}{$ 1.83 \pm 0.0201 $} & \cellcolor{gray!6}{\boldmath{$ 1.76 \pm 0.0351 $}} & \cellcolor{gray!6}{$ 110 $}\\
\hspace{1em}5 & \boldmath{$ 1730 \pm 73 $} & $ 2900 \pm 9.19 $ & \boldmath{$ 4.88 \pm 0.0242 $} & $ 4.93 \pm 0.02 $ & $ 270 $\\
\addlinespace[0.3em]
\multicolumn{6}{l}{\textbf{Dataset 6}}\\
\hspace{1em}\cellcolor{gray!6}{0.1} & \cellcolor{gray!6}{\boldmath{$ 1260 \pm 15.7 $}} & \cellcolor{gray!6}{$ 4220 \pm 25.5 $} & \cellcolor{gray!6}{\boldmath{$ 0.506 \pm 0.00908 $}} & \cellcolor{gray!6}{$ 0.758 \pm 0.000811 $} & \cellcolor{gray!6}{$ 12 $}\\
\hspace{1em}0.3 & \boldmath{$ 1240 \pm 58.2 $} & $ 4450 \pm 33.2 $ & \boldmath{$ 0.539 \pm 0.0335 $} & $ 0.769 \pm 0.000792 $ & $ 34 $\\
\hspace{1em}\cellcolor{gray!6}{0.5} & \cellcolor{gray!6}{\boldmath{$ 1240 \pm 14.3 $}} & \cellcolor{gray!6}{$ 4090 \pm 31.8 $} & \cellcolor{gray!6}{\boldmath{$ 0.601 \pm 0.00416 $}} & \cellcolor{gray!6}{$ 1.02 \pm 0.00611 $} & \cellcolor{gray!6}{$ 57 $}\\
\hspace{1em}0.7 & \boldmath{$ 1170 \pm 19.6 $} & $ 3920 \pm 632 $ & \boldmath{$ 0.762 \pm 0.00647 $} & $ 1.18 \pm 0.179 $ & $ 80 $\\
\hspace{1em}\cellcolor{gray!6}{0.9} & \cellcolor{gray!6}{\boldmath{$ 1340 \pm 75.6 $}} & \cellcolor{gray!6}{$ 3660 \pm 59.4 $} & \cellcolor{gray!6}{\boldmath{$ 0.828 \pm 0.0391 $}} & \cellcolor{gray!6}{$ 1.32 \pm 0.00631 $} & \cellcolor{gray!6}{$ 100 $}\\
\hspace{1em}1.1 & \boldmath{$ 1440 \pm 2.84 $} & $ 3900 \pm 23.3 $ & \boldmath{$ 0.961 \pm 0.0141 $} & $ 1.34 \pm 0.00494 $ & $ 130 $\\
\hspace{1em}\cellcolor{gray!6}{2} & \cellcolor{gray!6}{\boldmath{$ 2200 \pm 332 $}} & \cellcolor{gray!6}{$ 6610 \pm 13.4 $} & \cellcolor{gray!6}{$ 1.67 \pm 0.0175 $} & \cellcolor{gray!6}{\boldmath{$ 1.47 \pm 0.000724 $}} & \cellcolor{gray!6}{$ 230 $}\\
\hspace{1em}5 & \boldmath{$ 5540 \pm 60 $} & $ 11500 \pm 11.3 $ & $ 4.62 \pm 0.0131 $ & \boldmath{$ 4.49 \pm 0.016 $} & $ 570 $\\*
\end{longtable}
\endgroup{}

\paragraph{The adaptive ridge recovers the zones well}

\begin{figure}
\centering
\subfloat[True signal (areas not shown)]{
\includegraphics[width = 0.42\textwidth]{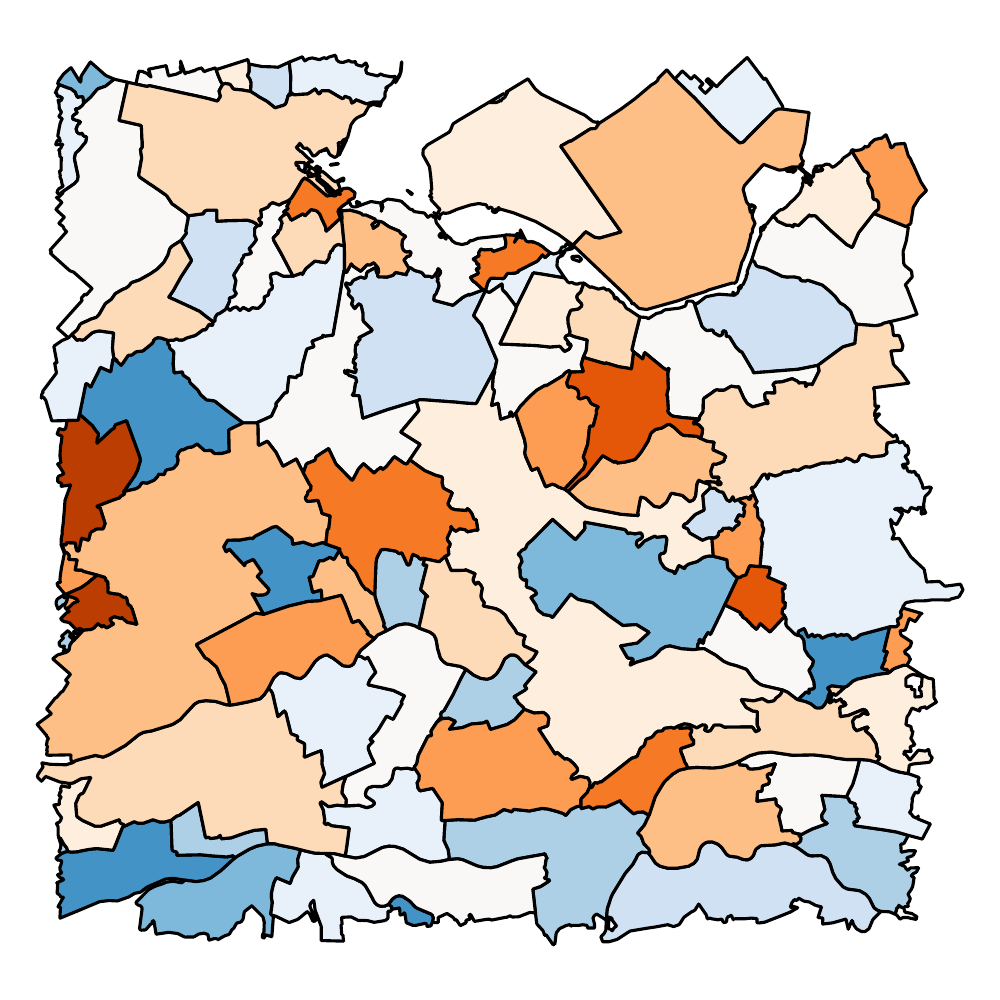}
\label{fig:utrecht_neigh_pc_municip_baseline_sigma5}
}
\hfil
\subfloat[Noisy signal]{
\includegraphics[width = 0.42\textwidth]{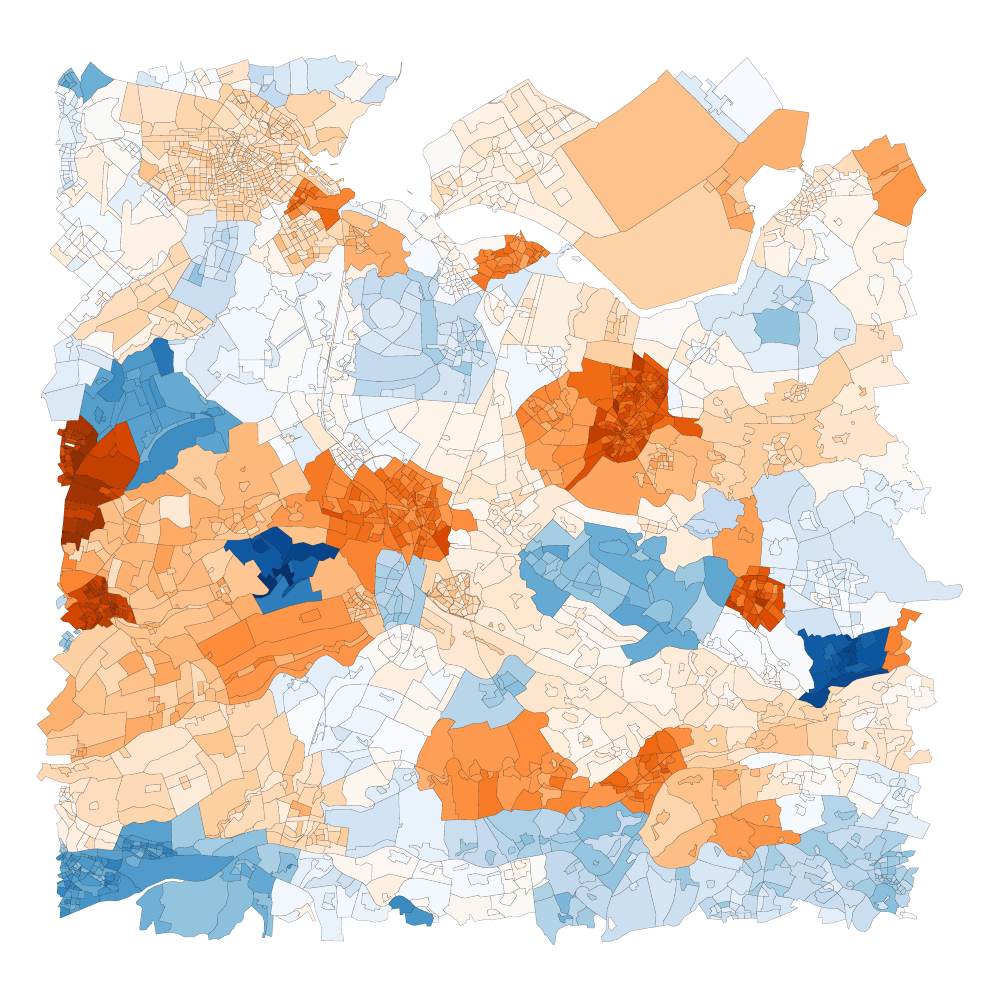}
\label{fig:utrecht_neigh_pc_municip_unseg_sigma5}
}
\hfil
\subfloat[Estimated signal -- \texttt{agraph}]{
\includegraphics[width = 0.42\textwidth]{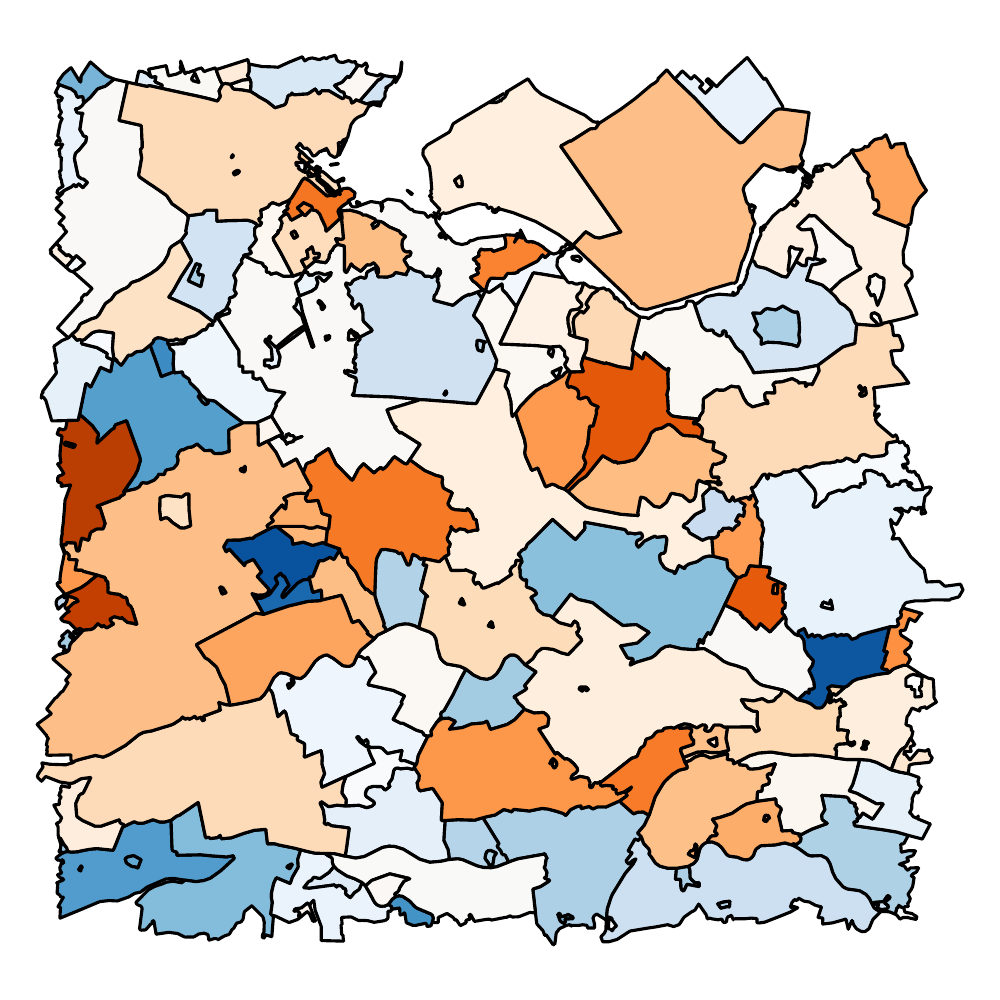}
\label{fig:utrecht_neigh_pc_municip_seg_sigma5}
}
\hfil
\subfloat[Difference between true and estimated -- \texttt{agraph}]{
\includegraphics[width = 0.42\textwidth]{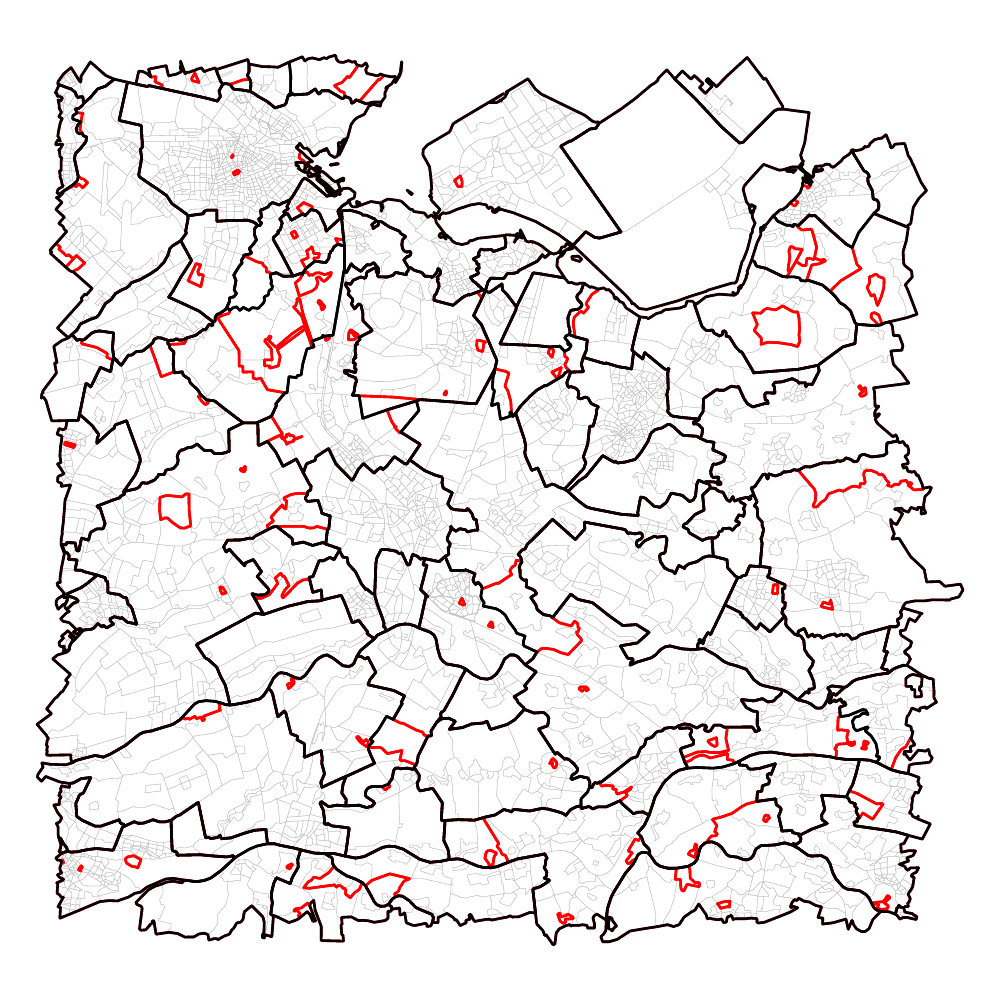}
\label{fig:utrecht_neigh_pc_municip_seg_overlay_sigma5}
}
\hfil
\subfloat[Estimated signal -- \texttt{flsa}]{
\includegraphics[width = 0.42\textwidth]{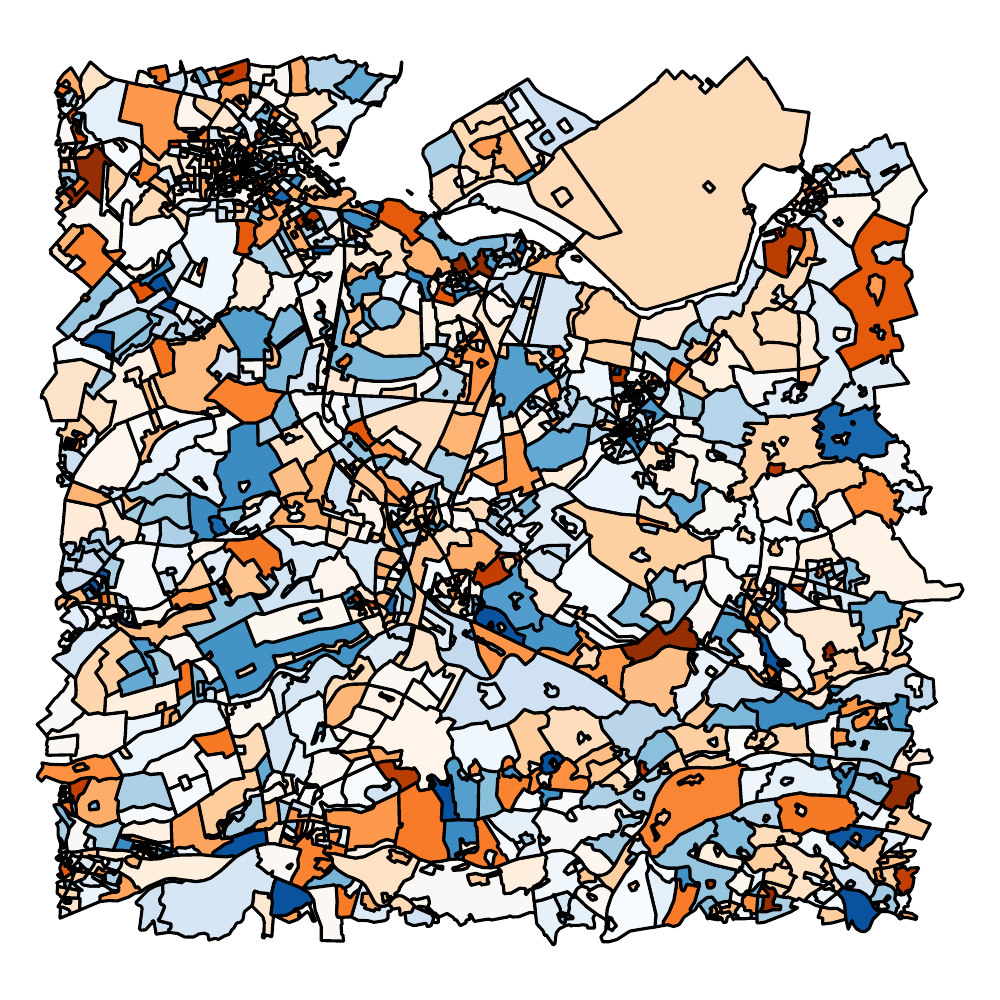}
\label{fig:utrecht_neigh_pc_municip_seg_sigma5}
}
\hfil
\subfloat[Difference between true and estimated -- \texttt{flsa}]{
\includegraphics[width = 0.42\textwidth]{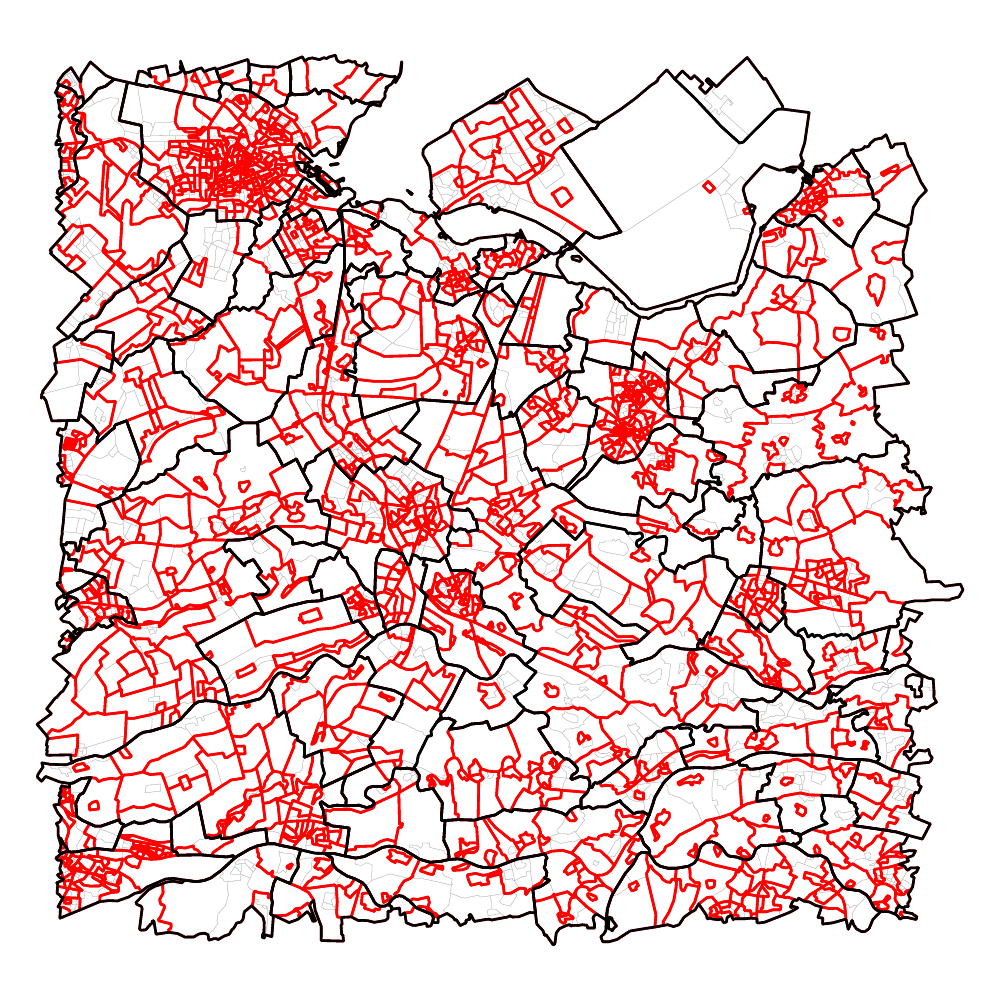}
\label{fig:utrecht_neigh_pc_municip_seg_overlay_sigma5}
}
\caption{Illustration of the piecewise constant estimate obtained by our method on Dataset 4. \emph{a)} True piecewise constant signal $\bm\theta$; \emph{b)} Raw signal $\bm{x}$ generated using $\sigma = 0.5$; \emph{c)} and \emph{e)} Estimated signal $\bm{\hat{\theta}}$ using the AIC; \emph{d)} and \emph{f)} Representation of the difference between the true (black) and estimated (red) signals. The panels share the same color scale.}
\label{fig:utrecht_neigh_pc_municip_plot}
\end{figure}

Our method performs well in terms of estimating~$\bm{q}$ (in terms of RMSE).
We now verify that it performs well at estimating the zones.

We can first assess this by visual inspection. Figure~\ref{fig:utrecht_neigh_pc_municip_plot} illustrates one estimate from Dataset 4, with $\sigma = 0.5$ and using the AIC. Out of the $99$ true zones, our method estimates $161$ different areas where flsa estimates $1231$. The overlay plot (Figures~\ref{fig:utrecht_neigh_pc_municip_plot}\emph{d)} and~\ref{fig:utrecht_neigh_pc_municip_plot}\emph{f)}) shows that the adaptive ridge estimate has a few more areas of small size, but recovers the correct shape of most areas. In comparison, the flsa estimates too many zones.

We can also quantify how well the estimated zones fit the true zones. Consider the problem of estimating zones as a problem of clustering, where the data points are the areas. Using this viewpoint, we use the Rand index \cite{rand1971objective} to measure how well each method estimates the true zones. We obtain a Rand index of $0.98$ for the adaptive ridge and $0.96$ for flsa and an ajusted Rand index of $0.85$ for our method and $0.07$ for flsa.

\paragraph{Comparing adaptive ridge and flsa runtime}

\begin{figure}[htb]
    \centering
    \includegraphics[width = 0.9\linewidth, keepaspectratio]{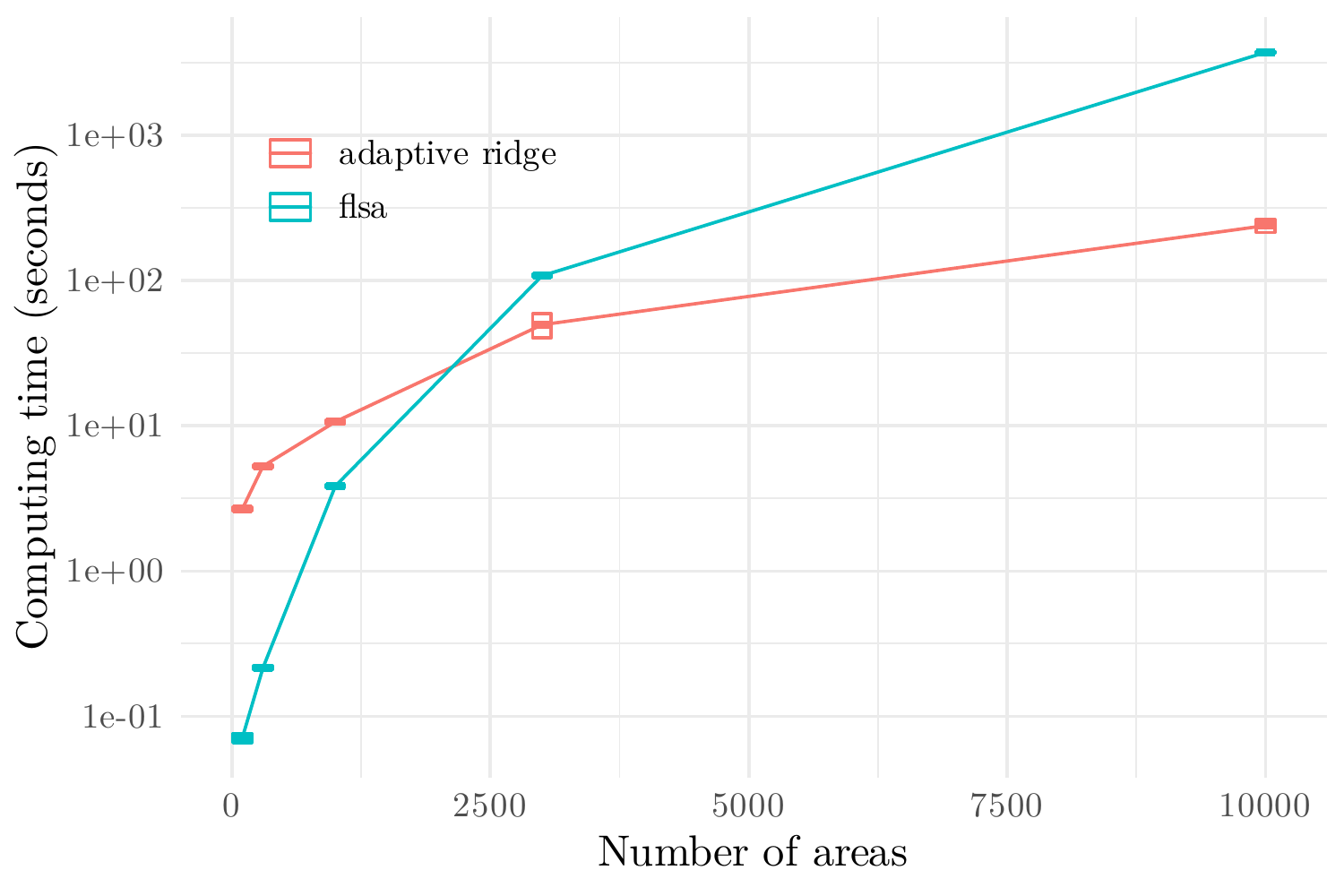}
    \caption{\label{fig:computation_time}Computing time in seconds of both methods for datasets of different sample sizes.}
\end{figure}

We want to compare the computation time of our method to that of flsa.
To that end we run both methods on planar graphs of different size. We use the adjacency graph of the neighborhoods on the whole of Netherlands ($p = 12920$, see Figure~\ref{fig:dataset6_zones}) and consider connected subgraphs by selecting a node and adding all vertices within a certain (graph-based) distance of that node.
We generate signals as iid normalized Gaussian samples and run both methods on a grid of $50$ values of $\lambda$.
The computing times are summarized in Figure~\ref{fig:computation_time}. The flsa is faster than the adaptive ridge for small graphs (less than $1000$ nodes) and the adaptive ridge is faster for large graphs (more than $3000$ nodes).

Runtime experiments (not show here) show that our method's runtime edge increases as the graph is more connected.

\subsection{Spatial segmentation of overweight in the Netherlands}

\begin{figure}
\centering
\subfloat[With $1021$ zones]{
\includegraphics[width = 0.45\textwidth]{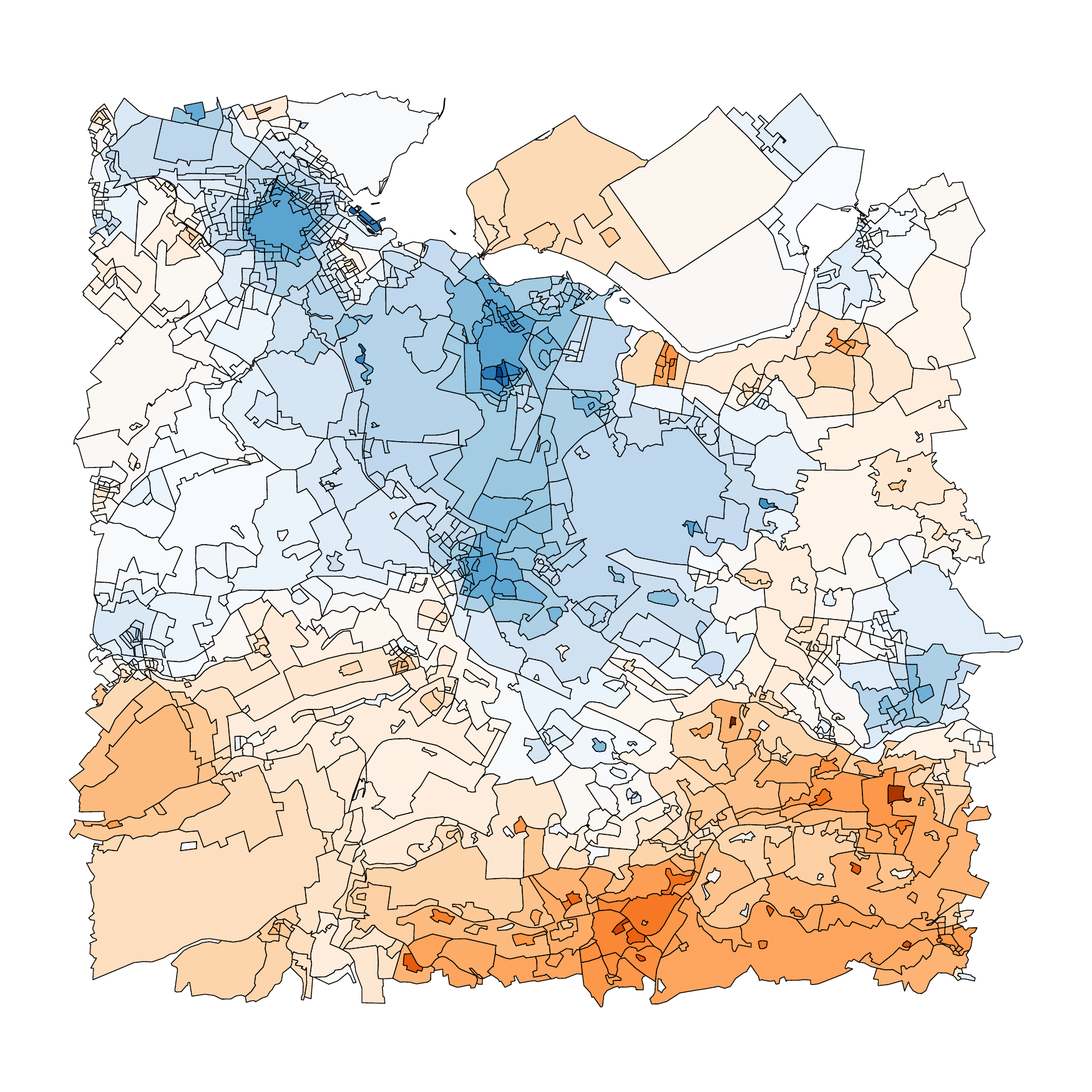}
\label{fig:mrf_cov_1}
}
\hfil
\subfloat[With $700$ zones]{
\includegraphics[width = 0.45\textwidth]{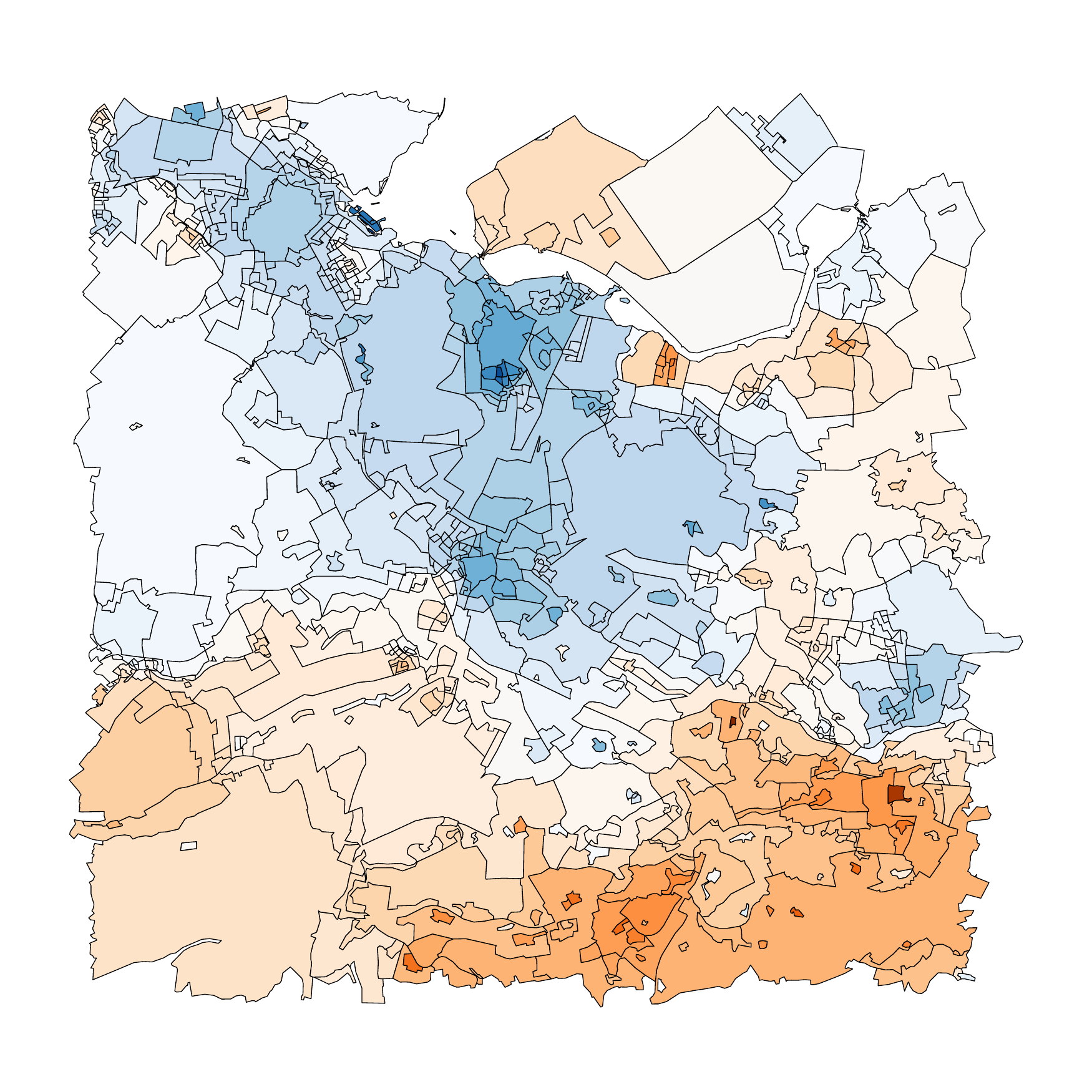}
\label{fig:mrf_cov_2}
}
\hfil
\subfloat[With $457$ zones]{
\includegraphics[width = 0.45\textwidth]{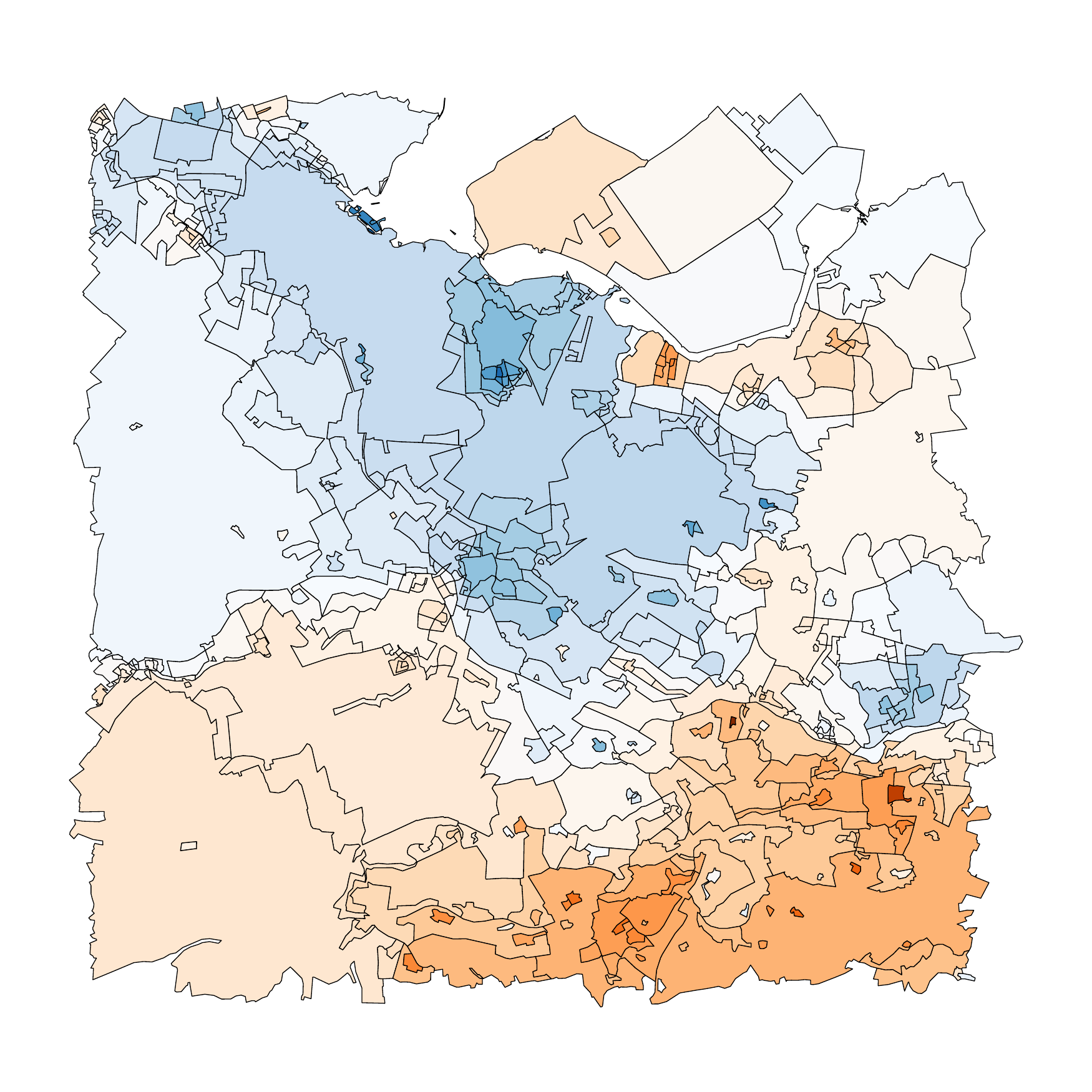}
\label{fig:mrf_cov_3}
}
\hfil
\subfloat[With $158$ zones]{
\includegraphics[width = 0.45\textwidth]{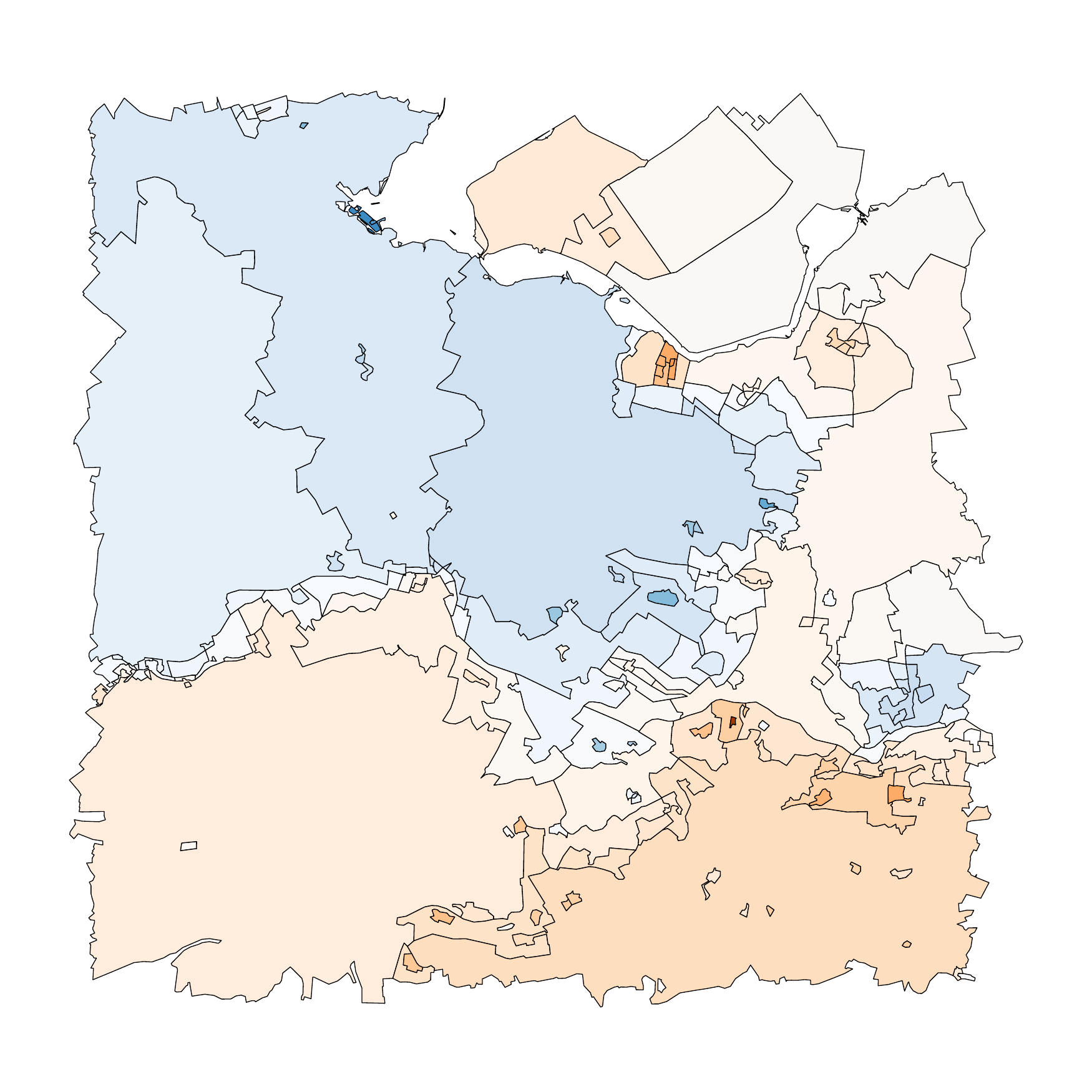}
\label{fig:mrf_cov_4}
}
\caption{Segmented spatial effect for overweight for different penalties. The results correspond to model dimensions of (a) $1020$, (b) $700.0$, (c) $457$, and (d) $158$ (corresponding to $1097$, $744$, $504$, and $163$ selected zones, respectively). The selected zones are delimited with thick lines.}
\label{fig:mrf_cov}
\end{figure}

Figure \ref{fig:mrf_cov} shows the result of the spatial segmentation of the log-odds ratios, i.e., the neighborhood effects, for the 2,955 neighborhoods in the Netherlands.
The models selected by the selection criteria are either underpenalized (model dimension of~$2558.3$ for GCV) or overpenalized (model dimensions of~$18.9$ for AIC and~$4.7$ for BIC).
Consequently, we display on Figure~\ref{fig:mrf_cov} the spatial segmentation for four different penalties, corresponding to model dimensions of (a) $1020$, (b) $700$, (c) $457$, and (d) $158$.

The fusion of areas is not happening uniformly. This can be seen for example in the south-west corner north of Dordrecht, where higher-than-expected dark region does not get fused with its neighbouring areas, while the background gets more and more segmented into larger zones with piecewise constant values.

Remarkably, large number of neighborhoods are selected as being different from their surrounding neighborhoods, even as the penalty gets very large. Typical examples is the IJburg island east of Amsterdam, the former fishers town Bunschoten-Spakenburg north of Amersfoort, and some small villages in the south.

\section{Discussion}

\subsection{Simulation study}

We have run our method on simulated data from 6 simulation settings, varied levels of noise standard deviation. We have compared it to the flsa, in terms of RMSE, model dimension, and ability to recover an accurate estimate of the zones.
Our method estimates signals of slightly lower RMSE for most noise levels (except for the high noise levels), but of greatly lower dimensions.
This comes from the fact that the adaptive ridge solves an~$L_{q}$ penalized problem, with~$q\to 0^{+}$ \citep[see][Section~S1]{goepp2021RegularizedBidimensionalEstimation}.
This is a desired property for obtaining zones that are easier to interpret. Moreover, the estimation of the zones, seen as a task of clustering the areas, is of better quality for our method.
Our method runs slower than flsa for small and medium number of areas ($\leq 1000$) but runs faster for large numbers of areas ($\geq 3000$).

\subsection{Real-data application}
To start with, for the real-data application, the identification of zones with higher or lower prevalence than expected was done in two steps: first generate data, then apply the segmentation to these data. This is done for two reasons. First, it is for computation reasons impossible to combine these two steps. The iterative nature of the adaptive ridge would require refitting the small area estimation model many times. One model fit takes a few minutes, so the total computation time would take hours or days. Second, the small area estimation model is fit on individual data. These data are only available in a secured environment hosted by Statistics Netherlands. Outside this environment, it only allowed to work with aggregated data.

With regard to the spatial segmentation, our method has identified zones that have higher or lower overweight prevalence than can be expected based on the demographic and socio-economic characteristics of neighborhoods alone. However, the total number of zones is still relatively large, here in the order of hundreds of zones. This is caused by neighborhoods that have a particular high or low value compared to their adjacent neighborhoods. From a practical point of view, one still has to visualize the results.

We have found that the BIC does not provide a satisfying selection: only one zone.
The question of which penalty parameter to favor depends on the how simple \emph{vs} intricate we want the \emph{blue zones} that are inferred by the model to be.
We note that different penalties provide different levels of information: \ref{fig:mrf_cov_1} simplifies the original data significantly while retaining much information about neighborhoods with specifically higher or lower prevalence than expected, while \ref{fig:mrf_cov_4} gives extended zones which are easier to interpret.

This collection of estimates with decreasing levels of granularity can serve policy makers as a tool to choose not only the optimal spatial distribution of the neighborhoods, but also the spatial scale to target their health-improving strategies. For example, a policy maker could focus either on the zone in the south-east (high odds ratio) when targeting on a large scale, or, on the other hand, focus on a specific neighborhood when targeting on a small scale.

We only considered overweight here. We also looked at other health-related indicators, like smoking, alcohol use and self-reported health. If blue zones would exist in the Netherlands, we would expect to see similar patterns of the spatial effect term. However, this was not the case. For example, the spatial term for overweight was negatively correlated with the spatial term for smoking (-0.24) and heavy drinking (-0.49). We found a "positive" correlation with self-reported health (0.41), i.e. a lower than expected overweight is positively associated with a better self-reported health. It is however outside the scope of this paper to explain these discrepancies.

The interpretability of the blue zones obtained by segmenting the overweight spatial term is somewhat limited. 
Considering the good performance of our method on simulated data, this does not question the quality of the method.
Rather, it hints that our method may be too simplistic for this application, as no other covariates are included inside the model.

\subsection{Methodological discussion \& conclusion}
This work introduces a method for graph-based signal segmentation applied to detecting piecewise constant effects in areal data.

The application is based on the assumptions that the areal discretization if adapted to the geographical distribution, in that not too much information about the distribution is lost when discretizing.
Moreover, the areal data is considered only through its adjacency structure. 
Thus, the segmentation obtained is of good quality if the graphical distance is close to the geographical distance.
This is usually the case for most administrative units, as the division into areas is fairly regular.

%


Besides having data values for each area, our method also requires a measure of precision of these values. This determines the weights that areas receive. If a value has been observed with a low precision, the corresponding area gets less weight and will be more quickly fused with its adjacent areas. It is also important to take the correlation (i.e. covariance or precision matrix) between areas into account. A positive correlation will oppose the penalization, while a negative correlation will facilitate the penalization. If no covariance information is available, one can still use a diagonal matrix with variances. If no variance information is available, one can use the identity matrix. In that case all areas receive equal weights.

For large graphs, our iterative method for segmentation over a graph is competitive with similar fused-type penalties. Its application to spatial data has shown to yield sparse models, which can make the spatial effect simpler to interpret.
Note that our method applies to disconnected, non-planar graphs, with possible applications to graph signal processing.

Our method also has some limitations. It makes the underlying assumption that the division into areas is regular enough, which is not always met in spatial statistics. Moreover, when~$\bm{\Sigma}$ is known, the method is computationally efficient under the assumption that $\bm\Sigma$ is sparse, which can limit its application to other types of problems.

Future works include extending the method to the case of linear regression:~$\bm{x} \sim \mathcal{N}(\bm{X}\bm{\theta}, \bm{\Sigma})$, where~$\bm{X}$ is the design matrix. Indeed, in this can, the penalized likelihood takes the form~$\frac{1}{2}(\bm x - \bm{X} \bm \theta)^\intercal \bm\Sigma^{-1} (\bm x - \bm{X} \bm \theta) + \frac{\lambda}{2} \bm\theta^\intercal \bm{K}^{(l)} \bm\theta$.
Consequently, our method can be extended to the linear model with graph-based penalty by replacing~\ref{eq:ar_weighted_ridge} with~$\bm\theta^{(l)} = (\bm{X}^{\intercal} \bm\Sigma^{-1} \bm{X}^{\intercal} + \lambda \bm{K}^{(l - 1)})^{-1}\bm{X}^{\intercal}\bm\Sigma ^{-1}\bm x$.

Further, the model can be extended to the general linear model, where the errors follow a distribution inside the exponential family. Indeed, the generalized linear model is estimated using the Iteratively Reweighted Least Squares (IRLS) procedure \citep[][Section 2.5]{McCullagh1989GeneralizedLinearModels}, which solves a reweighted ridge problem. Since the adaptive ridge is also based on iterations over a ridge problem, extension to the exponential family of distributions would consist in replacing the matrix $\bm\Sigma^{-1}$ in \eqref{eq:ar_weighted_ridge} with a diagonal matrix, the entries of which are update at each step $l$.

To summarize, we have presented a new method for spatial segmentation of areal data. It uses the adaptive ridge technique to penalize over the differences between adjacent areas. The method yields a segmented estimate of the spatial effect in a computationally efficient way. The model only requires areal data values and the adjacency structure as input. The method is shown to perform well, yielding estimates sparser than the lasso-based method we used for comparison. The method can assist policy makers with their health-improving strategies.
An implementation of our method is publicly available as an R package at \texttt{github.com/goepp/graphseg}.

\subsection*{Acknowledgements}
The authors would like to thank Pr.~Olivier~Bouaziz for his usefuls remarks which helped improve the quality of this paper.
This work was partially funded by the National Institute for Public Health and the Environment (RIVM) through its Strategic Research Programme (SPR) which contributes to solutions to societal challenges through interdisciplinary research and by supporting innovation and capacity building at RIVM. This work was also partially funded by the French Foundation for Medical Research ("Fondation pour la Recherche Médicale").

\subsection*{Declaration of competing interests.} The authors declare no competing interests.

\subsection*{Author contributions}
Conceptualization JvdK and VG; Data curation JvdK; Formal analysis VG and JvdK; Funding acquisition JvdK; Investigation VG; Methodology VG; Project administration VG; Resources VG and JvdK; Software VG; Supervision JvdK and VG; Validation JvdK and VG; Visualization VG; Roles/Writing -- original draft JvdK and VG; Writing -- review \& editing N/A


\newpage
\appendix
\section{Convergence criterion for the algorithm}
\label{sec:appendix_convergence}

Define the weighted differences
\begin{equation*}
    \delta_{j,k}^{(l)} \triangleq v_{j,k}^{(l)} (\theta_{j}^{(l)} - \theta_{q}^{(l)})^{2} = \frac{(\theta_{j}^{(l)} - \theta_{q}^{(l)})^{2}}{(\theta_{j}^{(l)} - \theta_{q}^{(l)})^{2} + \varepsilon}.
\end{equation*}
With $\varepsilon$ very small, as the sequence $\bm\theta^{(l)}$ converges, the $\delta_{j,k}^{(l)}$s tend to either zero if the two values are close to equal, or one if the two values are different. Consequently, this quantity serves to diagnose the convergence of the algorithm.

More precisely, the stopping criteria is when the absolute differences between two consecutive values $\delta_{j,k}$ are smaller than a fixed tolerance, for all pairs $j\sim k$. This tolerance parameter is set to $10^{-8}$ is our implementation. We chose a very small tolerance to make sure that the algorithm has converge, but we advise taking a higher tolerance (e.g., $10^{-6}$) makes the estimating procedure run faster (for illustration, the real data application runs on a Intel Core i7 CPU in $1,130$ seconds with a tolerance of $10^{-8}$, versus $625$ seconds with a tolerance of $10^{-6}$).

At convergence, the estimated model is not sparse. Therefore we use a cutoff of 0.99 to round the $\delta_{j,k}^{(l)}$s to zero or one. The cutoff is purposefully set to a high value, which ensures that we only remove vertices between adjacent neighborhoods with very different spatial effects. Thus, our method is conservative in separating two neighborhoods.

Adjacent areas with a weighted difference of zero have been estimated to have the same spatial effect. This yields a partition of the areas into a set of connected subgraphs who each have been estimated to have the same underlying spatial effect. 
These subgraphs are the estimated zones.

Note that the algorithm developped here induces shrinkage of~$\bm{\hat{\theta}}$.
This is in contrast with the initial implementation of \cite{Frommlet2016AdaptiveRidgeProcedure}, which uses a two-step approach, using the adaptive ridge to estimate the zones, and then estimating the effect on each zones using unpenalized estimation.
Our approach using shrinkage was giving better results on simulations (results not shown here).

\end{document}